\documentclass[longnamesfirst,apjl]{emulateapj}
\usepackage{apjfonts,natbib,epsfig}
\tighten

\newcommand{\tableskip}{\\[-6pt]}      
\newlength{\tskip}
\newlength{\colwidth}

\newcommand{\beq}{\begin{equation}}      
\newcommand{\eeq}{\end{equation}}      
\newcommand{\beqa}{\begin{eqnarray}}      
\newcommand{\eeqa}{\end{eqnarray}}      

\newcommand{\bn}{\hat{\bf n}}

\newcommand{\rad}{r}    

\begin{document}      
\title{Clustering of the IR Background Light with Spitzer: Contribution from Resolved Sources}      
\author{Ian Sullivan\altaffilmark{1}, Asantha Cooray\altaffilmark{2}, Ranga-Ram Chary\altaffilmark{1},      
James~J.~Bock\altaffilmark{1,3}, Mark Brodwin\altaffilmark{3}, 
Michael~J.~I.~Brown\altaffilmark{4},
Arjun Dey\altaffilmark{5},Mark Dickinson\altaffilmark{5}, Peter Eisenhardt\altaffilmark{3},    
Henry C. Ferguson\altaffilmark{6}, Mauro Giavalisco\altaffilmark{6}, 
Brian Keating\altaffilmark{7}, Andrew Lange\altaffilmark{1},     
Bahram Mobasher\altaffilmark{6},
 William T. Reach\altaffilmark{8}, Daniel Stern\altaffilmark{3},
Edward L. Wright\altaffilmark{9}}      
\altaffiltext{1}{Division of Physics, Mathematics, and Astronomy, California Institute of Technology, Pasadena, CA 91125}      
\altaffiltext{2}{Center for Cosmology, Department of Physics and Astronomy, University of California, Irvine, CA 92697. E-mail: acooray@uci.edu}      
\altaffiltext{3}{Jet Propulsion Laboratory, 4800 Oak Grove Drive, Pasadena, CA 91109}      
\altaffiltext{4}{Princeton University Observatory, Peyton Hall, Princeton, NJ 08544}
\altaffiltext{5}{National Optical Astronomy Observatory, 950 N. Cherry Ave., Tucson, AZ 85719}    
\altaffiltext{6}{Space Telescope Science Institute, 3700 San Martin Dr., Baltimore, MD 21218}    
\altaffiltext{7}{Department of Physics, University of California, La Jolla, CA}
 \altaffiltext{8}{Spitzer Science Center, California Institute of Technology, Pasadena, CA 91125}      
 \altaffiltext{9}{UCLA Astronomy, PO Box 951562, Los Angeles, CA 90095-1562}

\begin{abstract}      
      
We describe the angular power spectrum of resolved sources      
at 3.6 $\mu$m ($L$-band) in {\it Spitzer} imaging data of the GOODS HDF-N, the GOODS CDF-S, and the NDWFS Bo\"otes field in several source magnitude bins.     We also measure angular power spectra of resolved sources in the Bo\"otes field at $K_S$ and $J$-bands    
using ground-based IR imaging data. In the three bands, $J$, $K_S$, and $L$, we     detect the clustering of galaxies on top of the shot-noise power spectrum at multipoles between      
 $\ell \sim 10^2$ and $10^5$. The angular power spectra range from the large, linear scales to small, non-linear scales of galaxy clustering,  and in some magnitude ranges,  show departure from a power-law clustering spectrum. We consider a halo model to describe clustering measurements and  to establish the halo occupation number parameters of IR bright galaxies at redshifts around one. The typical halo mass scale at which    
 two or more IR galaxies with $L$-band Vega magnitude between 17 and 19 are found in the same halo is    
between $9 \times 10^{11}$ M$_{\sun}$ and $7 \times 10^{12}$ M$_{\sun}$ at the 1$\sigma$ confidence level; 
this is consistent with the previous halo mass estimates  for bright, red galaxies at $z\sim1$. We also    
extend our clustering results  and completeness-corrected faint source number counts in GOODS fields to
understand the underlying nature of unresolved sources responsible for IR background   (IRB) anisotropies that were
detected in deep {\it Spitzer} images.  While these unresolved  fluctuations were measured at sub-arcminute angular scales,    
if a high-redshift diffuse component associated with first galaxies exists in the IRB, then it's clustering properties are 
best studied with shallow, wide-field images     
that allow a measurement of the clustering spectrum from a few degrees to arcminute angular scales.
\end{abstract}      
      
      
\section{Introduction}      
      
The intensity of the cosmic near-infrared background (IRB) is a measure of the total light emitted by stars and galaxies in the      
Universe. While the absolute background has been estimated by space-based      
experiments, such as the  Diffuse Infrared Background Experiment (DIRBE; Hauser \& Dwek 2001) and       
the Infra-Red Telescope in Space (IRTS; Matsumoto et al. 2005), the total IRB intensity measured      
still remains fully unaccounted for by sources:       
only 13.5 $\pm$ 4.2 nW m$^{-2}$ sr$^{-1}$ is resolved to point sources at 1.25 $\mu$m (Cambr\'esy et al. 2001), while current direct measurements range from 25-70 nW m$^{-2}$ sr$^{-1}$.       
At wavelengths greater than 3 $\mu$m, with a       
total IRB intensity of 12.4 $\pm$ 3.2 nW m$^{-2}$ sr$^{-1}$ (Wright \& Reese 2000) at 3.6 $\mu$m,      
the ``missing source'' problem is less significant with 5.4 nW m$^{-2}$ sr$^{-1}$      
resolved to point sources in {\it Spitzer} data (Fazio et al. 2004a; Franceschini et al. 2006), while fluctuation analyses of {\it Spitzer} data yield 10.6$^{+0.63}_{-1.95}$ nW m$^{-2}$ sr$^{-1}$ (Savage \&  Oliver 2005).       We refer the reader to Kashlinsky (2005) 
for a recent review on the IRB including a summary of past attempts to understand the intensity excess relative to the
background predicted by resolved source number counts.
      
Primordial galaxies at redshifts 8 and higher, especially those involving Population III stars, are generally invoked       
to explain the missing IR flux between 1 $\mu$m and 2 $\mu$m, with most of the intensity associated with redshifted Lyman-$\alpha$ emission during reionization      
(e.g., Santos, Bromm \& Kamionkowski 2002; Salvaterra \& Ferrara 2003; Cooray \& Yoshida 2004; Fernandez \& Komatsu 2006).       
While models of high-redshift Pop III populations can explain the ``missing'' IRB, these models       
run into several difficulties if such sources were to account for all of the missing IR intensity. These include the      
high efficiency required to convert baryons to stars in first galaxies (e.g., Madau \& Silk 2005)      
and limits from deep IR imaging data that suggest a lack of a large population of       
high-redshift dropouts (Salvaterra \& Ferrara 2006).      
The recently revised optical depth to reionization (Page et al. 2006),       
with reionization around a redshift of 10      
instead of the previous estimates of 20, decreases the number density of first sources that
are required for reionization. In return, the implied fractional contribution from first sources
to the total intensity of the IRB  is lowered.
    
Still, one does expect some contribution to the IRB from sources that reionized the Universe, though      
the exact intensity of the IRB from such sources is yet unknown both theoretically and observationally.    
As pointed out in Cooray et al. (2004; also, Kashlinsky et al. 2004), if a high-redshift population contributes significantly       
to the IRB, then these sources are expected to leave a distinct signal in the       
anisotropy fluctuations of the near-IR intensity, when compared to the anisotropy spectrum associated with low-redshift      
sources. Using results from a fluctuation analysis in deep {\it Spitzer} imaging data      
with resolved point sources removed from the image down to a deeper magnitude than previous studies on this topic,      
 Kashlinsky et al. (2005)      
claimed a potential detection of the clustering signature of high-redshift sources at wavelengths of 3.6, 4.5, and 5.8 $\mu$m.      
A previous attempt to understand the nature of this excess clustering in
{\it Spitzer} data, when resolved sources are removed down to a magnitude level of 22.5 in the $L$-band,     suggested that
it could be the clustering signature of galaxies at redshifts greater than 5, with a total contribution to the IRB at the level of    
$\sim$ 1 to 2 nW m$^{-2}$ sr$^{-1}$ in  the $L$-band (Salvaterra et al. 2006); at shorter wavelengths, the intensity of this     
background remains uncertain as the spectrum of this 
excess fluctuation component is only established with {\it Spitzer} for $\lambda \geq$ 3.6 $\mu$m.  While this suggestion was simply based on
a model description of the fluctuation spectrum, whether such a  scenario is consistent with faint source counts in deep {\it Spitzer}
images is yet to be considered.
      
If the 1 to 2 $\mu$m IRB is not fully resolved by extragalactic sources,  then it could be that most of the missing source problem      
is due to deficiencies in accounting for the zodiacal light (Dwek, Arendt \& Krennrich 2005).      
This possibility may be further consistent with some suggestions that,       
subject to a priori assumptions on their intrinsic spectral shape, the observed blazar TeV spectra rule      
out a significant IRB from extragalactic sources (e.g., Aharonian et al. 2006).      While the
point source photometry may underestimate the total IR flux in each resolved source by missing the outer regions when constructing source catalogs,
it is unlikely that this effect alone can explain the factor of 2 to 3 difference between predicted and measured IRB intensity given that
point source photometry is unlikely to miss more than 5\% to 10\% of the total flux (see Totani et al. 2001 for a discussion).
Such a small residual, however, could produce an excess in fluctuations of the IRB, 
with the angular spectrum tracing that of the bright resolved sources.

To separate various possibilities related to the nature of intensity as well as anisotropy excesses, 
further analyses of near-IR background data is required.       
If the excess background is due to missing flux in low-z galaxies, then one would expect fluctuations in the background      
to strongly correlate with resolved point sources.      
The contribution from zodiacal light is known only to within an upper limit, but if residual zodiacal light is the reason, then spatial clustering of the IRB will peak at degree or more angular scales due to its smoothness.       
Thus, to distinguish between these possibilities, it is necessary to further understand
spatial clustering at IRB wavelengths. Here, we concentrate on the      
fluctuations in IRB due to resolved sources. We study both the shape      
and the amplitude of the angular power spectrum and use these measurements to make a relative comparison between       
resolved and unresolved IRB clustering, with the latter from measurements in  Kashlinsky et al. (2005).      
We also connect our clustering measurements with faint number counts and an underlying halo model of the IR
galaxy distribution within the large-scale structure to study if faint galaxies could be responsible for the
IRB fluctuations seen with {\it Spitzer}.
      
In this paper, we make clustering measurements at 3.6 $\mu$m in      
the Great Observatories Origins Deep Survey (GOODS; Dickinson et al. 2003) HDF-N field, 
the GOODS CDF-S field,  and the NOAO Deep Wide-Field Survey (NDWFS; Jannuzi \& Dey 1999) 
Bo\"otes field with the source sample divided into      
several magnitude bins in each of these catalogs. 
The data come from {\it Spitzer} IRAC imaging of these fields. We refer the reader
to Dickinson et al. (in preparation) for IRAC observations of the GOODS field and Eisenhardt et al. (2004; also see
Fazio et al. 2004a) for the Bo\"otes-field IRAC shallow survey. In addition to {\it Spitzer} images,
we also measure clustering of resolved sources in the Bo\"otes field in ground-based images
at $K_S$ and $J$-bands with data taken by the FLAMINGOS Extragalactic Survey (FLAMEX; Elston et al. 2006). 

Here we present clustering measurements in terms of an angular power spectrum in
 multipole space. Due to simplicity in the measurement, 
galaxy clustering is also measured in real space in terms of the angular
correlation function or $w(\theta)$. As discussed in Tegmark et al. (2002), Fourier or multipole-space measurements
should be preferred over the real space due to the simple behavior of the covariance matrix and the fact that
multipole-space measurements can be easily related to the underlying three-dimensional power spectrum without a complex
window function that maps fluctuations in Fourier space to that of the angular space. 
In the case of the real space correlation function,  clustering at different angular scales are also correlated
(Eisenstein \& Zaldarriaga 2001) and this also lead to a complex covariance matrix that must be accounted for
when model fitting the measurements. 

Despite these considerations, real-space clustering measurements are easy to make and have been studied in the literature
at 3.6 $\mu$m and other {\it Spitzer} IRAC bands (Oliver et al. 2004;
Fang et al. 2004).  While model fitting was considered in terms of a power-law, a clear departure from a power-law model fit is present in 
angular clustering of 3.6 $\mu$m galaxies in the   {\it Spitzer} Wide-Area Infrared Extragalactic (SWIRE) as described by Oliver et al. (2004;
 for example, see their Figure~1 for clustering in Lockman and EN1 fields that indicates
a rise above a power-law at small angular scales). On the other hand, clustering measurements in the First Look  Survey (FLS)
at 3.6 $\mu$m by Fang et al. (2004) do not show clear evidences for a departure from the power-law (see, Figure~1 in Fang et al. 2004).
This could be due to the smaller number of galaxies used for clustering measurements, when compared to statistics in Oliver et al. (2004).
Here, using the Bo\"otes-field IRAC shallow survey over 6 deg.$^2$, we find a departure from the power-law description and model
that departure in terms of the halo model for galaxy clustering (Cooray \& Sheth 2002). A departure from a power-law
description, however, is not surprising for 
galaxy clustering at high redshifts due to non-linear clustering at angular scales less than a few arcminutes (Ouchi et al. 2005; Lee et al. 2005).     
Such departures have been detected for Lyman-break galaxies at $z \sim 3$ to 5 (Ouchi et al. 2005; Lee et al. 2005; Cooray \& Ouchi 2006),
 as well as at low redshifts    with surveys such as SDSS (Zehavi et al. 2005), DEEP2 (Coil et al. 2004), and COMBO-17 (Phleps et al. 2004).
These departures were expected, and are described, in terms of analytical approaches such as 
the halo model (Cooray \& Sheth 2002) and conditional luminosity functions (Cooray  2006a; van den Bosch et al. 2006).     
      
The {\it Paper} is organized as following: in the next section, we briefly summarize our   data and in \S~3 we describe statistics in our
catalogs. In \S~4, we describe the procedure we used to measure clustering of IR sources and
in \S~5 we present  a summary of the analytical model used to describe clustering.      
Based on model fits, we extract quantities related to how IR galaxies occupy dark matter halos. In \S~6, we      
discuss our results in the context of recent IRB anisotropy measurements with {\it Spitzer} and future wide-field
surveys with {\it Akari} (Matsuhara et al. 2006), Cosmic Infrared Background Explorer (CIBER; Bock et al. 2006), and {\it Spitzer} 
   to resolve the ``missing source'' problem in the near-IR background intensity.      
We conclude with a summary in \S~7.   Throughout this paper, we take a reference cosmology model with parameters $\Omega_m=0.3$,
$\Omega_\Lambda =0.7$, $h=0.7$ (unless left as a free parameter, such as in the luminosity function), and a normalization to
the matter power spectrum at 8 h$^{-1}$ Mpc scales $\sigma_8=0.84$.
                  
\section{Imaging Data}      
      
To generate the angular power spectrum, we make use of
the source catalogs in the GOODS and the NDWFS-Bo\"otes fields that were extracted using the SExtractor
program (Bertin \& Arnouts 1996) using 3.6'' and 5'' apertures,
respectively. Vega magnitudes were calculated from the catalog flux
values by converting to AB magnitudes and adding the appropriate
offset from Kashlinsky (2005), and throughout this paper magnitudes refer to
Vega magnitudes. We refer the reader to Table~1 for basic statistics related to our fields such as the number of
galaxies in specific magnitude ranges and the contribution to IRB from those galaxies when the counts are integrated.
We now provide a brief summary of each of the two fields.

\subsection{Bo\"otes field}

We make use of both {\it Spitzer} IRAC imaging data as well as the ground-based    
$J$- and $K_S$-band imaging data of the Bo\"otes field from the FLAMEX survey (Elston et al. 2006)
where the  source catalogs are publicly available \footnote{http://flamingos.astro.ufl.edu/extragalactic/overview.html}.   
We refer the reader to Eisenhardt et al. (2004) for details
related to this IRAC Shallow survey and the first results from these data,
while Fazio et al. (2004a) contains a discussion of number counts.
The {\it Spitzer} Bo\"otes field covers a total area close to 8.5 deg.$^2$, but due to non-uniform coverage and to avoid
introducing a complicated window function for clustering measurements in the multipole space, we restrict the discussion to
the central 6.3 deg.$^2$ here. While ground-based FLAMEX imaging data of the Bo\"otes field 
cover a total area of almost 7 deg.$^2$, again due to gaps in the data,
we concentrate on 3.9 deg.$^2$ with uniform coverage at the center of the field.

To understand the extent to which Bo\"otes field imaging is complete, we performed standard simulations by placing ten thousand artificial point sources
randomly across the Bo\"otes mosaic in each 0.2 magnitude bin.  We perform
object detection and extraction  exactly as we had carried out with the original survey 
catalog as described in Eisenhardt et al. (2004). We quantify completeness as the
recovered fraction of point sources in each magnitude bin.
The formal 50\% completeness limit of the catalog  occurs at a 3.6$\mu$m magnitude of 19.3.  To the
3.6$\mu$m limit of 17.8 mag, the IRAC Shallow survey is over 85\% complete. Due to completeness and to avoid systematic biases from
stellar sources in the galaxy that dominate at magnitudes below 15 (see, Fazio et al. 2004a) our clustering measurements in the
{\it Spitzer} Bo\"otes field  is restricted to galaxies in the magnitude range between 17 and 19. In the case of
ground-based surveys, the 50\% completeness limit for 5 $\sigma$ detection of point sources
is in $K_S$ band is  19.5 (Elston et al. 2006). Thus, we restrict clustering
from ground-based images to the magnitude range between 17 and 19 in $K_S$ band. Since typical galaxy color $J-K_s$ is around
1.5 (Elston et al. 2006), we can extend clustering studies in J-band to magnitudes around 21. In the case
of {\it Spitzer} $L$-band galaxy sample, typical color $L-K$ is about 1.1 (see, also Oliver et al. 2004).

In Figure~1 right panel, we show the completeness corrected counts in L-band and those published by Fazio et al. (2004a). 
We find a good agreement with completeness corrected counts and those published by Fazio et al. (2004a) using the same imaging
data.  We do not show counts in J and $K_S$ bands as these are discussed in detail in Elston et al. (2006).

\subsection{GOODS fields}

Imaging data of GOODS fields (both CDF-S and HDF-N) with {\it Spitzer} IRAC cover roughly 0.046 deg.$^2$.
In both fields,  catalogs of sources in the GOODS IRAC 3.6$\mu$m mosaics were generated using the SExtractor \citep{Ber96} software. 
A 9$\times$9 Mexican-hat kernel was used with a FWHM of 2.4$\arcsec$ (Dickinson et al. 2006 for further details).
Monte-Carlo simulations were performed on the IRAC images wherein 
artificial sources, convolved with the {\it Spitzer} point spread function, 
were input at random positions on to the mosaic. At faint flux densities, 
almost all sources are point sources at the IRAC spatial resolution of 
$1.6\arcsec$ FWHM (Fazio et al. 2004b). 100 sources were added in each iteration, which is 
negligibly small compared to the $\sim$ 14,000 sources which are in the IRAC 
catalog. The flux distribution of artificial sources was flat in 
$\log(S_{3.6})$  enabling equal numbers of sources in each logarithmic flux 
bin. The 50
completeness corrections outlined in the next paragraph, we extend clustering measurements  down to a magnitude limit of 22.5.

Sources in the original GOODS catalogs were first matched to sources in
the catalog generated on the image with the fake sources, with a 1$\arcsec$ 
matching threshold. If the sources in the original catalog were not matched 
it implies that the flux and position were both affected by 
confusion/blending from an artificial source. The positions of the input artificial 
sources are known and these are matched with the unmatched sources in the 
catalog with the same positional tolerance.
An artificial source of a particular flux would then be detected if it is within 1$\arcsec$ of 
its original position although its extracted flux might be significantly different from the 
input flux if it landed close to an existing bright IRAC source. The artificial source would, however, 
be undetected if its position in the catalog is more than an arcsecond from the original position.
The flux distribution of input and output sources is transformed to a matrix P$_{ij}$ where index $i$ 
is the input flux and index $j$ is the output flux (Chary et al. 2004). 
The nature of the P$_{ij}$ matrix is such that for a particular $i$, the sum over all $j$ is less than unity. 
This is the completeness correction factor that we use for the GOODS CDF-S and HDF-N catalogs.

\begin{figure*}[!t]
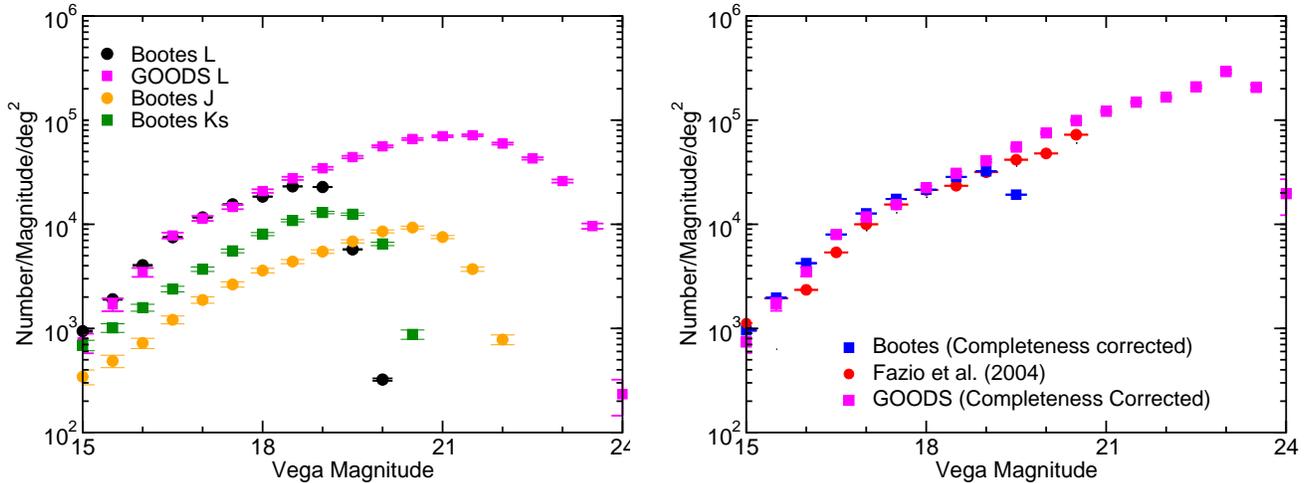
      
\centerline{\psfig{file=Fig1a.eps,width=3.3in,angle=0}    
\hspace{0.3cm}
\psfig{file=Fig1b.eps,width=3.3in,angle=0}}    
\caption{{\it Left}: Number counts in the GOODS HDF-N and Bo\"otes fields with {\it Spitzer} IRAC at $L$-band, and Bo\"otes fields from the ground at $J$- and $K_S$-bands using data from the FLAMEX survey, as described in \S~2.         
{\it Right}: Comparison to Fazio et al. (2004a)  counts and the completeness corrected counts of the Bo\"otes field used in the
analysis of this paper. Between the magnitude range of 17 and 19 used for clustering measurements in the Bo\"otes field,
we find a good agreement with counts extracted for the present analysis and the ones studied by Fazio et al. (2004a) for
total background intensity measurement at 3.6 $\mu$m. Note that Fazio et al. (2004a) counts shown here are the completeness  corrected
ones at magnitudes fainter than 16. 
}
\label{fig:num}      
\end{figure*}

\begin{figure}[!t]      
\centerline{\psfig{file=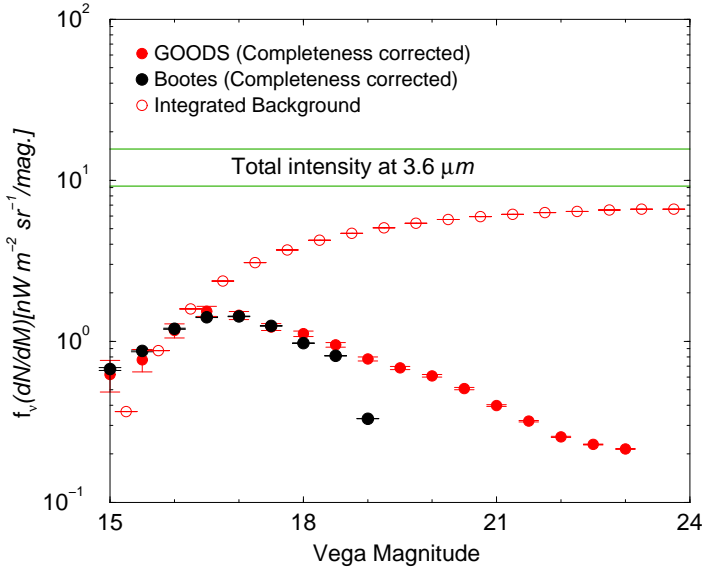,width=3.0in,angle=-90}}
\caption{The fractional (solid symbols; in nW m$^{-2}$ sr$^{-1}$ magnitude$^{-1}$) and total (open symbols; in
nW m$^{-2}$ sr$^{-1}$) contribution to the intensity of the IRB at 3.6 $\mu$m from
galaxy counts in GOODS and Bo\"otes  fields. The two horizontal lines show the measured range of the total intensity at 3.6
$\mu$m of 12.4 $\pm$ 3.2 nW m$^{-2}$ sr$^{-1}$ (Wright \& Reese 2000). The counts are completeness corrected.
Using GOODS counts, we measure a total intensity for the IRB at 3.6 $\mu$m of $\sim$ 6.6 nW m$^{-2}$ sr$^{-1}$ consistent with previous estimates.}
\label{fig:numirb}      
\end{figure}

\begin{figure*}[!t]      
\centerline{
\psfig{file=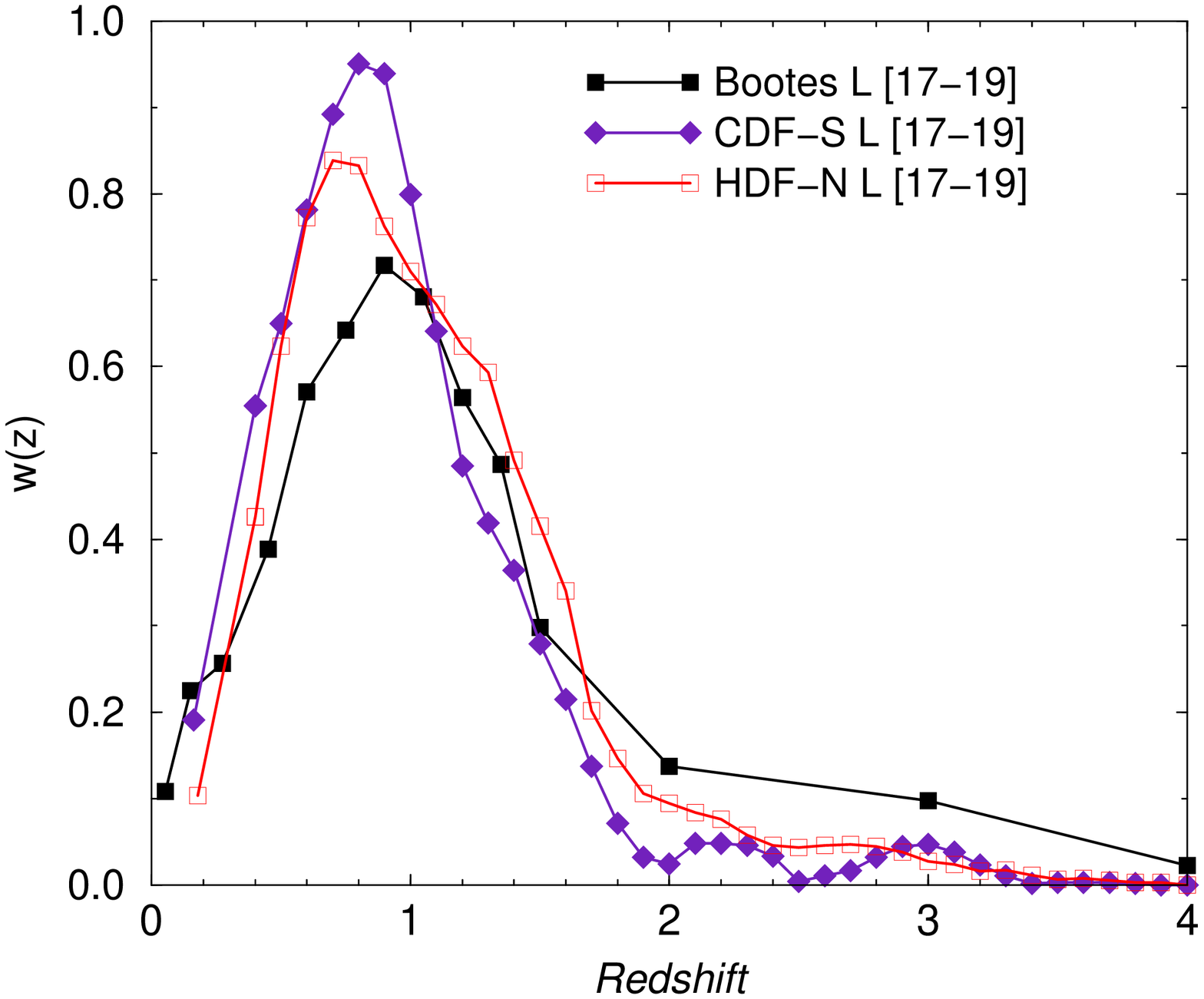,width=2.8in,angle=-90}
\psfig{file=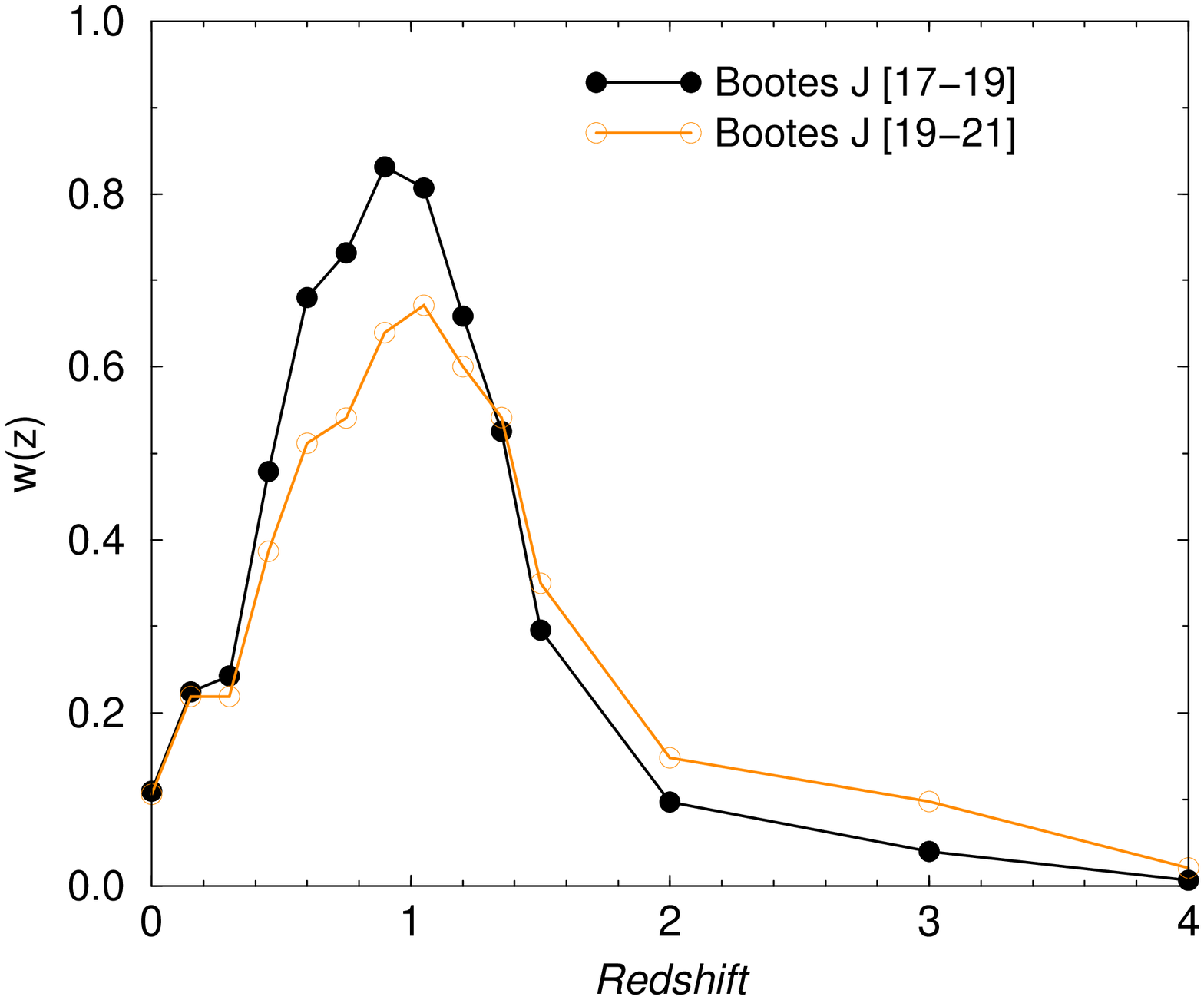,width=2.8in,angle=-90}}
\caption{{\it Left}: 
The normalized ($\int dz w(z)=1$) redshift distribution of $L$-band galaxies with magnitudes between 17 and 19 as estimated photometrically    
for the GOODS HDF-N and CDF-S fields (Mobasher et al. 2004) as well as for the Bo\"otes field (Brodwin et al. 2006).
 We take into account these  distributions when model fitting the $L$-band clustering measurement in this    
magnitude bin.
{\it Right}: The redshift distribution of $J$-band galaxies with magnitudes between 17 and 19 and between 19 and 21, as estimated again based on
photometric redshifts.}
\label{fig:zdist}      
\end{figure*}

\section{Statistical Properties of GOODS and Bo\"otes fields}

Here, we make use of two types of surveys: the Bo\"otes field provides us with a shallow, 
wide-field image to measure clustering out to 
degree angular scales for bright sources (down to    
a magnitude limit of 19 in the $L$-band), while the GOODS HDF-N and the GOODS CDF-S {\it Spitzer} images allow     
us to measure clustering of resolved sources at 10 arcminute angular scales and below
down to a very deep magnitude limit.  In combination, the Bo\"otes image allows us to measure clustering from degree scales to sub-arcminute scales
for bright galaxies and GOODS catalogs allow clustering measurements of resolved sources down to the faintest level of 22.5 in $L$-band.
This is at the magnitude level at which resolved sources were removed from the fluctuation analysis in Kashlinsky et al. (2005).
Thus, clustering studies with the two fields, in return,
allow us to establish the confusion level when making measurements related to the unresolved    
diffuse component and to understand if the fluctuations detected by Kashlinsky et al. (2005) can be explained as
simply extrapolating the faint number counts to below the point-source detection threshold applied in Kashlinsky et al. (2005).

\subsection{Number counts and the IRB intensity}

In Figure~1 left and right panels, we show the original number counts and completeness corrected number counts from various catalogs,
respectively. In the right panel, the completeness corrected counts are compared with those
published by Fazio et al. (2004a). 
As shown in Fig.~1 (right panel) and for completeness corrected counts in both Bo\"otes and GOODS fields, we 
find good agreement to the extent that we can compare the fields reliably. In Fig.~2 we show the fractional and integrated contribution to the 
total intensity of the IRB in each of the magnitude bins (e.g., Totani et al. 2001).   
For comparison, we also show the total intensity of the IRB at 3.6 $\mu$m from Wright \& Reese (2000).

Integrating the source counts corrected for the stellar number counts, $\int S \, dN/dS \, dS$, extragalactic sources
between the magnitude range of 8 and 17 produce a cumulative background of 2.4 nW m$^{-2}$ sr$^{-1}$, 
while incomplete source counts between $L$-band magnitudes of
17 and 19 generate a IRB intensity of about 1.8
nW m$^{-2}$ sr$^{-1}$. The total intensity of Bo\"otes source counts between 17 and 19 magnitudes increases to 2.1 nW m$^{-2}$ sr$^{-1}$ when counts are 
corrected for completeness.  In combination, this is roughly 37\% of the total background at 3.6 $\mu$m of 
about 12 nW m$^{-2}$ sr$^{-1}$ (Wright \& Reese 2000). 

For sources fainter than 19th magnitude,
the background varies depending on the completeness corrections and the assumed slope below the point-source 
detectability level with the GOODS images. For completeness corrected source counts between $L$-band magnitudes of 19 and 22.5, we
find a total background of 2.3 nW m$^{-2}$ sr$^{-1}$. If the faint-end GOODS slope in Fig.~1 left panel is extrapolated to
a magnitude limit well below 30, we find a cumulative background of 1.2 nW m$^{-2}$ sr$^{-1}$ for sources fainter than a magnitude of 22.5.
As we discuss later in Section~4.2, such a slope is inconsistent with fluctuation measurements of the unresolved
background by Kashlinsky et al. (2005), though such a slope is suggested in the study of the one-point distribution function
of the unresolved flux by  Savage \& Oliver (2005). The faint-end slope that describes the anisotropy measurements
of Kashlinsky et al. (2005) suggests a background intensity from sources below a magnitude limit of 22.5 of
0.08 nW m$^{-2}$ sr$^{-1}$. Compared to the total intensity of about  12 nW m$^{-2}$ sr$^{-1}$, we find that
resolved source counts and unresolved background clustering explain about 55\% of the background for extragalactic
sources, which is consistent with the resolved fraction in Fazio et al. (2004a), but lower than  the resolved
fraction suggested in Savage \& Oliver (2005).

\subsection{Redshift distribution}

The photometric redshift distribution for galaxies with magnitudes between 17 and 19 in 3.6$\mu$m
is computed as described in Brodwin et al. (2006).  We show this redshift distribution in Figure~3 left panel
with a comparison to GOODS photometric redshift distributions in the same magnitude range for both HDF-N
and CDF-S fields.  The GOODS distributions are described in Mobasher et al. (2004) and are
 accurate to $(z_{\rm phot}-z_{\rm spec})/(1+z_{\rm spec})\sim0.05$ statistically.
There is a broad agreement between Bo\"otes and GOODS distributions with the distributions peaking 
at redshifts between 0.8 and 0.9.
Based on numerical calculations, we determined that the fractional contribution to the clustering power spectrum
from galaxies at redshifts greater than 2 with magnitudes between 17 and 19 in $L$-band is below a few percent 
at the angular scales probed by Bo\"otes and GOODS catalogs. Thus,
with a clear peak of the redshift distribution at $z \sim 0.9$, we can easily interpret parameters extracted 
from galaxy clustering  as those related to sources at this redshift instead of accounting for the
corrections coming from the high-redshift tail of the redshift distribution.  

Furthermore, while not exact in terms of
selection criteria or observing wavelengths,  the redshift distribution for the bright IR galaxy sample between 17 and 19 magnitudes in
the $L$-band
is similar to the galaxy redshift distribution in the high-redshift end of the DEEP2 survey. Thus, our statistical results on
quantities such as the halo mass hosting IR galaxies can be compared with those established based on the DEEP2 clustering measurements
(Coil et al. 2004).  In the right panel of Figure~3 we show the
 redshift distributions of $J$-band galaxies with magnitudes between 17 and 19 and between 19 and 21, as estimated photometrically.
These again peak at $z \sim 1$ and we use them when model fitting the clustering measurements in the $J$-band with source catalogs
generated by the FLAMEX survey.

\section{Clustering Measurement Technique}

The measurement of galaxy clustering in the multipole space is now well described in the literature with a wide range of applications with
data from several large-scale structure surveys. These techniques borrow from ideas that have been developed for clustering measurements in
the multipole space in cosmic microwave background (CMB) anisotropy maps.
In general, the available techniques fall into two approaches. First, starting from a technique described in Peebles (1973),
clustering can be derived either through  spherical harmonic estimation 
with moments calculated by summing over all pixels.
The second approach is to apply a maximum likelihood method to data to estimate the power spectrum directly, especially when
data involve large numbers of pixels where direct computation of spherical transforms is computationally time consuming
 (e.g., Tegmark et al. 2002). Here, we make use of the former approach given the small area of our surveys and the ability 
to directly estimate the transform. 

To begin, in each catalog, the source count distributions were fit to a fine grid
of pixel size 1.8'' for Bo\"otes and 0.7'' for GOODS, with each pixel $i$ in
the grid containing the sum of the flux of all the sources contained within it, $I_i(\bn)$, where $\bn$ is the two-dimensional
angular vector on the sky. In comparison to usual clustering measurements of the galaxy distribution,
$I_i(\bn)$ is not just the number of galaxies in each pixel, but rather the total IRB intensity in that pixel given the
magnitude of  that galaxy.  Note that we put all flux within the pixel at the source center and do not
account for the large IRAC PSF. While this procedure ignores the spatial structure of individual sources, since we are measuring
large-scale clustering between sources rather than small-scale clustering at spatial scales below a typical source size, this
is not a concern for us for this study. It would, however, be a concern when studying clustering in the unresolved background (Section~6.2).

In summary, to extract $C_l$, we expand the $I_i(\bn)$ as
\begin{equation}
I_i(\bn) = \sum_{l=0}^{l_{\rm max}} \sum_{m=-l}^{m=l} I_{lm} Y_{lm}(\bn) \, ,
\end{equation}
where $Y_{lm}$ are the spherical harmonic functions. In principle, if data exists over the whole sphere, angular power spectrum is simply
\begin{equation}
C_l= \frac{1}{2l+1} \sum_{m=-l}^{l} \left[|I_{lm}|^2 - C_l^{\rm shot-noise}\right] \, ,
\end{equation}
where the factor $(2l+1)$ comes from the fact that at each multipole $l$, there are $(2l+1)$ independent modes.
In above, $C_l^{\rm shot-noise}$ arises from the discrete behavior of the two dimensional intensity field (see below) and acts as a source of
noise for clustering measurements with a finite density of sources (Peebles 1973) and this shot-noise can be written
as $C_l^{\rm shot-noise}=\int S^2 dN/dS\, dS$, where $dN/dS$ is the surface density of sources at flux $S$,
assuming all counts are extracted with no confusion.

In practice, however, one must correct clustering measurements for the incomplete sky coverage through an 
appropriate renormalization of the harmonics. Following Peebles (1973), for partial sky coverage, we can write an
estimate for the angular power spectrum as
\begin{equation}
 \hat{C}_l = \frac{1}{2l+1} \sum_{m=-l}^{l} \left[\frac{|I_{lm} - C_l^{\rm shot-noise} A_{lm}|^2}{B_{lm}} -  C_l^{\rm shot-noise} \right]\, ,
\end{equation}
where the two functions $A_{lm}$ and $B_{lm}$ are $A_{lm} = \int Y_{lm}(\bn) d\bn$ and  $B_{lm}  = \int |Y_{lm}(\bn)|^2 d\bn$,
respectively, and the  angular integrals are 
over the area of the survey.  While in the all-sky case, estimates of $C_l$ at each multipole $l$ is independent  of
each other, with partial sky coverage, adjacent estimates are correlated. Thus, one cannot represent the power spectrum with estimates
at each $\ell$. To decrease the correlations, we average adjacent $\hat{C}_l$ estimates over bins $\Delta l$ in the multipole space
such that $\Delta l$ is taken to be logarithmically dependent on central multipole or $\Delta = x l$, with $x$ is fixed at the level of 0.2.
With such large bins, using the window function in multipole space, $W_l = \sum_{m=-l}^{l} |A_{lm}|^2/(2l+1)$, we have established that the
power spectrum estimates are correlated at a level below a few percent and we neglect these correlations in our model fitting analysis.
While not pursued here due to simple survey geometry and the small area used for clustering measurements,
the above procedure can be optimized further to quickly calculate harmonic transforms over large areas on the sky using
specially designed pixelization schemes that have been developed for CMB  anisotropy studies (e.g., HEALPix;  Gorski et al. 2005)
and  pseudo-$\hat{C}_l$ estimates (e.g., Wandelt et al. 2001). 

As described earlier,
to avoid complicating the clustering measurement with images that have large gaps in the source distribution and a complex
window function $W_l$, we chose a smaller, central area within the Bo\"otes field    
of 6.3 square degrees with IRAC catalogs and a 3.9 square degrees field in ground-based images. Such a selection also guarantees that our    
clustering measurements are not affected by problems at the edge of the field and that we do not need to
decorrelate the measurements for the complex geometry. For the pixel sizes used, based on a Monte Carlo simulation, 
we have established that the contribution to the power spectrum from overlapping
sources was negligible.  However, our grid has empty pixels due to masking of bright sources or no data (e.g., muxbleed contamination)
which we describe through a mask and such empty pixels cannot be avoided since they occur throughout the field. 
To avoid introducing complications to the power spectrum measurement, we filled those random empty pixels with
white noise before estimating the power spectrum, where the white noise was estimated using the same procedure as the  boot-strap approach 
outlined below to establish errors for binned $C_l$. To test the extent to which such a procedure can affect our
clustering measurements, we utilized a Monte-Carlo approach by systematically masking out different regions that were not contaminated
or, instead of adding simple white noise distributed as a Gaussian, replacing the  contaminated regions with Poisson noise.
We found that clustering measurements do not vary beyond the error indicated and that large-scale clustering is unaffected by
contaminated pixels.

As outlined above our clustering measurements lead to an estimation of the total power spectrum,      
which includes both the shot-noise component, associated with the Poisson-noise or discreetness of the source counts, and the 
clustering      
component, associated with the true spatial distribution. The latter is the component that is of interest to
large-scale structure studies and, as outlined above, one must remove the shot-noise component.
There are two approaches to estimate the shot-noise. First, one can simply measure the total power spectrum including the shot-noise
and then account for the shot-noise part of the spectrum, $\hat{C}_l^{\rm shot-noise}$,      
by measuring the amplitude of clustering, that will scale simply as 
$\ell^2$ when plotted in terms of $l^2\hat{C}_l/2\pi$ or as a constant is simply considered as $\hat{C}_l$, at large $\ell$ values.
This assumes that at large $\ell$ values or small angular scales, clustering is dominated by pure noise. 
Since noise for our measurement is simply the shot-noise, 
this amplitude can be taken to be the shot-noise and that constant value can be 
removed from the total spectrum $\hat{C}_l$. The second approach is to use
the information related to existing number counts and simply do the integral  $\int S^2 dN/dS\, dS$  to 
estimate the expected shot-noise, since this is also a correct description of the shot-noise. 
In practice, we have found that the shot-noise component is      
more accurately measured using the amplitude of the total power spectrum at
smallest angular scales probed by the data just before the finite size of pixels become important than
simply evaluating  the integral over the number counts in the magnitude range for which clustering measurements are made.
This is due to the incompleteness in our measured number counts as we now describe.

To understand this, our clustering spectrum is estimated from the IRB intensity captured by individual sources with the image pixelized
in terms of flux. Incompleteness in the point source detection catalog leads to a systematic underestimate
of the flux in certain pixels. In addition to incompleteness, our clustering spectrum
is also affected by stellar sources that were not removed in the original catalog. 
In this case we may be overestimating the clustering amplitude from contamination related to stellar flux.
Thus, in addition to incompleteness of the sky coverage which is easy to correct for once the geometry of the survey area is specified, 
we use a two-step approach to correct for the angular power spectrum of galaxies in each of the magnitude bins.
First, our clustering measurements are corrected to account for incompleteness in the catalogs.
Using the difference between  $\int S^2 dN/dS\, dS$ for completeness
corrected counts and the one estimated directly from clustering data that is an underestimate of the total shot-noise,
we correct the shot-noise and clustering power-spectra as
\begin{eqnarray}
\tilde C_l^{\rm shot-noise} &=&  \hat{C}_l^{\rm shot-noise} \frac{\int S^2 dN/dS^{\rm complete}\, dS}{\int S^2 dN/dS^{\rm incomplete}\, dS} \, \nonumber \\
\tilde C_l &=& \hat{C}_l \left[\frac{\int S dN/dS^{\rm complete}\, dS}{\int S dN/dS^{\rm incomplete}\, dS}\right]^2 \, .
\end{eqnarray}
When making these corrections to the overall amplitude we are assuming that incompleteness 
affects the field uniformly  and that incompleteness correction has no angular structure. In optical surveys, incompleteness can be related 
to extinction and for large are surveys such as SDSS that span over 1000 deg.$^2$ or more, extinction varies across the field
significantly and these angular variations must be accounted for. For the present study involving IR galaxy clustering at angular scales of a 
few degree and below, such complex corrections can be safely ignored.

The completeness corrected counts in Figure~1 left panel results in a 18\% increase in the shot-noise,
which if not accounted for leads to an underestimate of power at small angular scales.
At arcminute scales where clustering measurements from the
Bo\"otes field overlap with the GOODS, we find good agreement between the measurements from these two fields when
Bo\"otes clustering measurements are corrected for completeness issues. While it is not necessary to
correct for completeness in the GOODS fields between magnitudes of 17 and 19 in the $L$-band,
for clustering measurements in fainter bins, we make use of the completeness calculations to correct for the
shot-noise and the amplitude.

Since we did not correct the catalogs for potential contamination from stars,
the shot-noise contains an additional contribution from the stellar counts in addition to galaxies whose clustering
we are attempting to measure. To correct for the shot-noise associated with stars, we make use of the model of Fazio et al. (2004a)
and calculate the shot-noise associated with stars in the magnitude range of 17 to 19. The tabulated counts in Fazio et al.
(2004a), based on two independent deep fields, lead to two different estimates of the stellar shot-noise correction. 
The stellar shot-noise is 6\% and 10\% of the original total shot-noise for the two fields and is a small, though non-negligible,
correction to the angular power spectrum. We show both cases when presenting our clustering measurements to highlight the small
difference, which is only important at arcminute and smaller angular scales. Note that the corrections related to stellar sources are
done similar to that of the catalog incompleteness and again we assume that confusion from stars are uniform throughout the image.

\begin{figure*}[!t]      
\centerline{\psfig{file=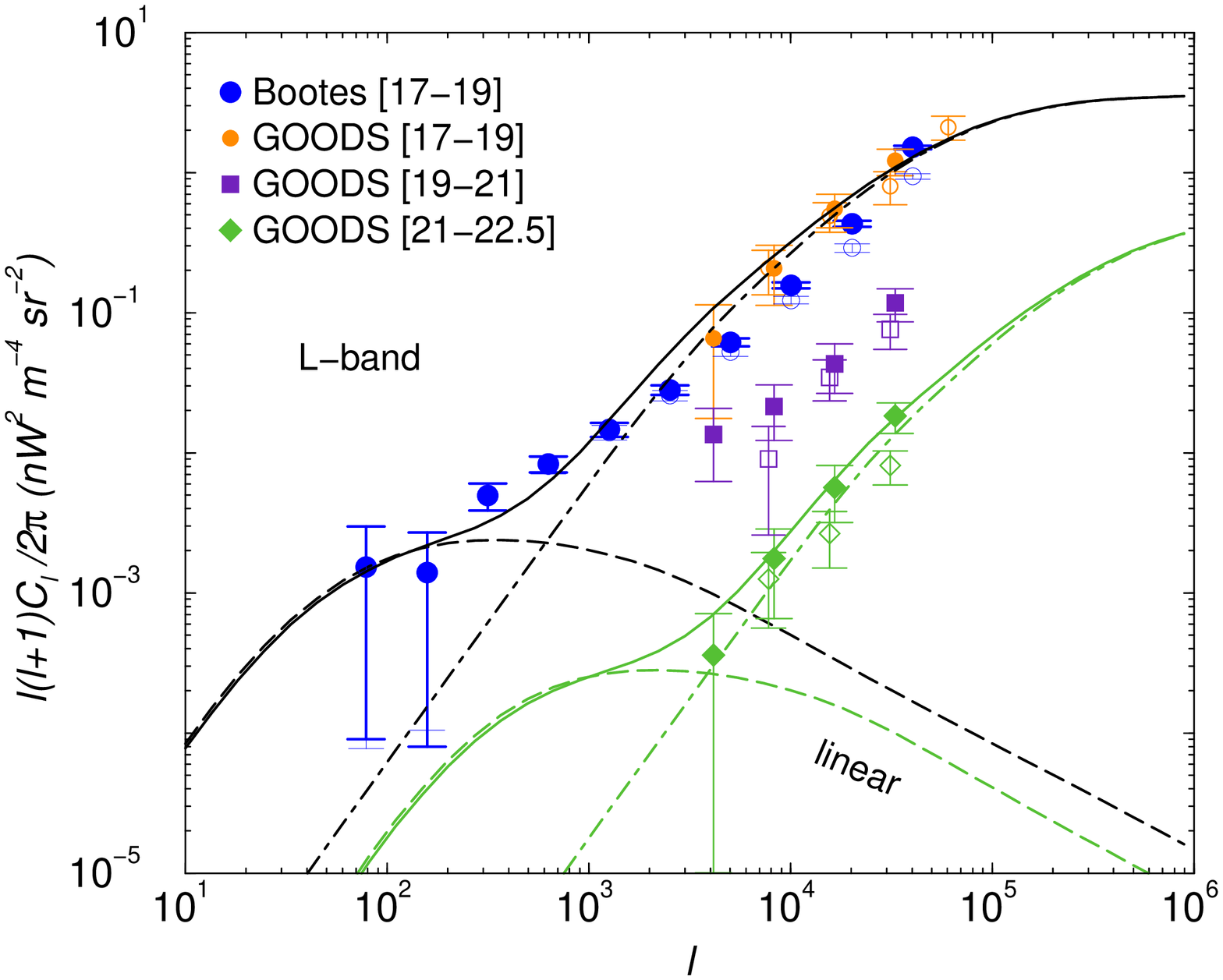,width=2.8in,angle=-90}    
\psfig{file=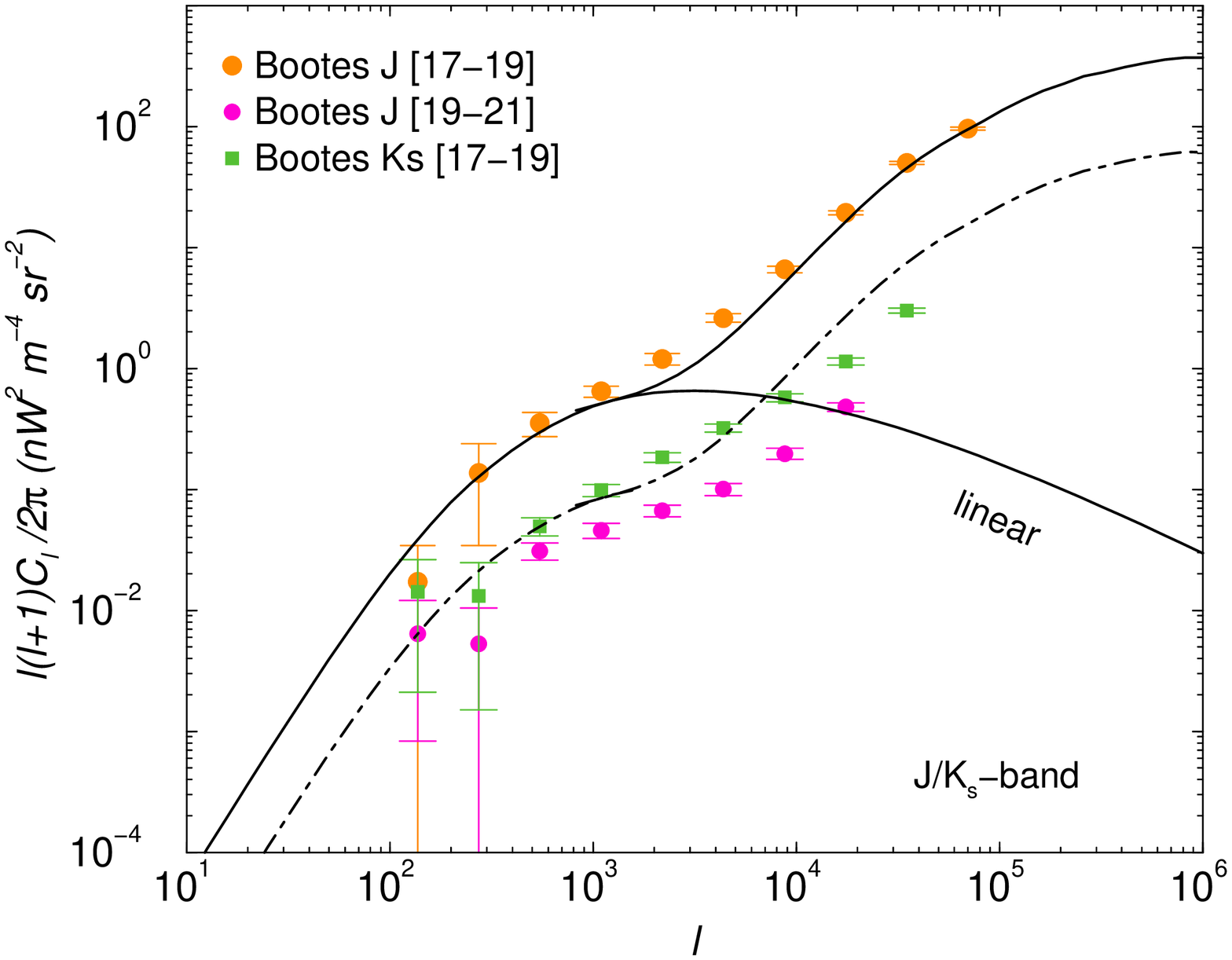,width=2.8in,angle=-90}}      
\caption{{\it Left:} $L$-band 
clustering measurements in both the GOODS fields and the NDWFS Bo\"otes field with {\it Spitzer} IRAC.     
The measurements are divided into magnitude bins in the catalog from      
17 to 19 (45 to 7 $\mu$Jy) (for both Bo\"otes and GOODS; top measurements), 19 to 21 (7 to 1.1 $\mu$Jy) (GOODS; middle), and 21 to 22.5 (1.1 to 0.28 $\mu$Jy) (GOODS; bottom).  The angular scale of clustering is related to the multipole $\ell$ through roughly $\theta \sim \pi/\ell$.
In the case of GOODS, filled symbols are the clustering measurements in the HDF-N while open symbols are from the CDF-S.    
The Bo\"otes clustering spectrum is extracted based on the completeness corrected shot-noise following counts in Figure~1 left panel. The open and filled symbols associated with the Bo\"otes clustering spectrum show the difference between
two estimates for the stellar contribution to the number counts and the shot-noise. This contribution is 6\% 
of the total shot-noise in the magnitude range of 17 to 19 in $L$-band for the open symbols and 10\% 
for the filled symbols, and this difference is relatively insignificant. 
The lines show model descriptions of the angular power spectrum of    
source clustering using CLF models in 17 to 19 and 21 to 22.5 magnitude bins (see Section~5.2)
and we show the total (solid lines), 1-halo term (dot-dashed lines), and the prediction under linear theory for galaxy clustering (dashed lines).
{\it Right:}  Clustering of resolved point sources in the ground-based imaging data of the Bo\"otes field    
at $J$ and $K_S$  bands in magnitude bins. 
Note the factor of 100 difference in the y-axis scale between the left and the right panels.     
This difference is consistent with the mean intensity of the    
IR background between these two different wavelengths. The curves here show the total (solid and dot-dashed for J and Ks-band clustering between 17 and 19 magnitudes, respectively), and for reference, in the case of J-band clustering, the expected linear model prediction.}
\label{fig:sloan}      
\end{figure*}

\begin{figure*}[!t]      
\centerline{\psfig{file=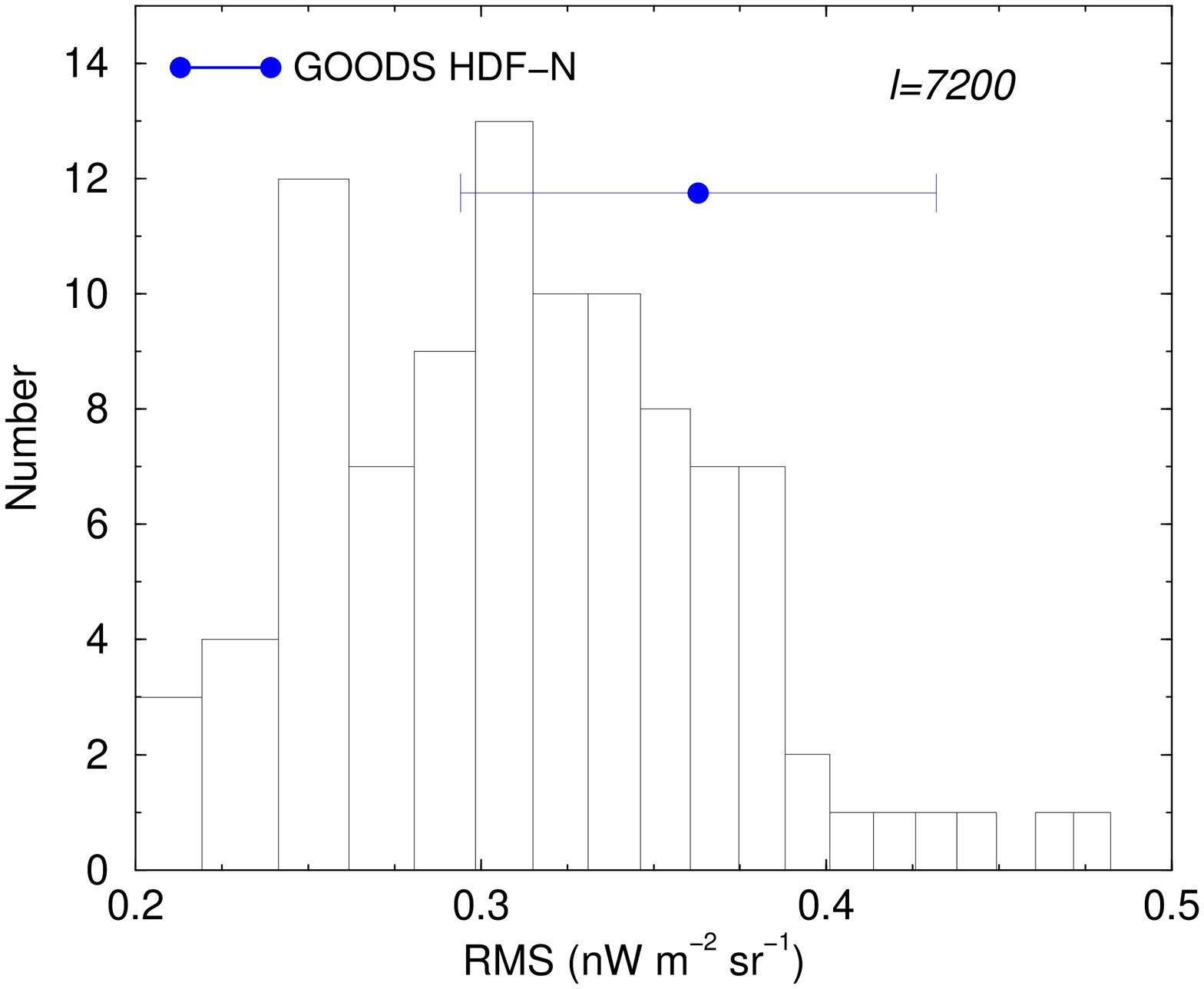,width=2.8in,angle=-90}    
\psfig{file=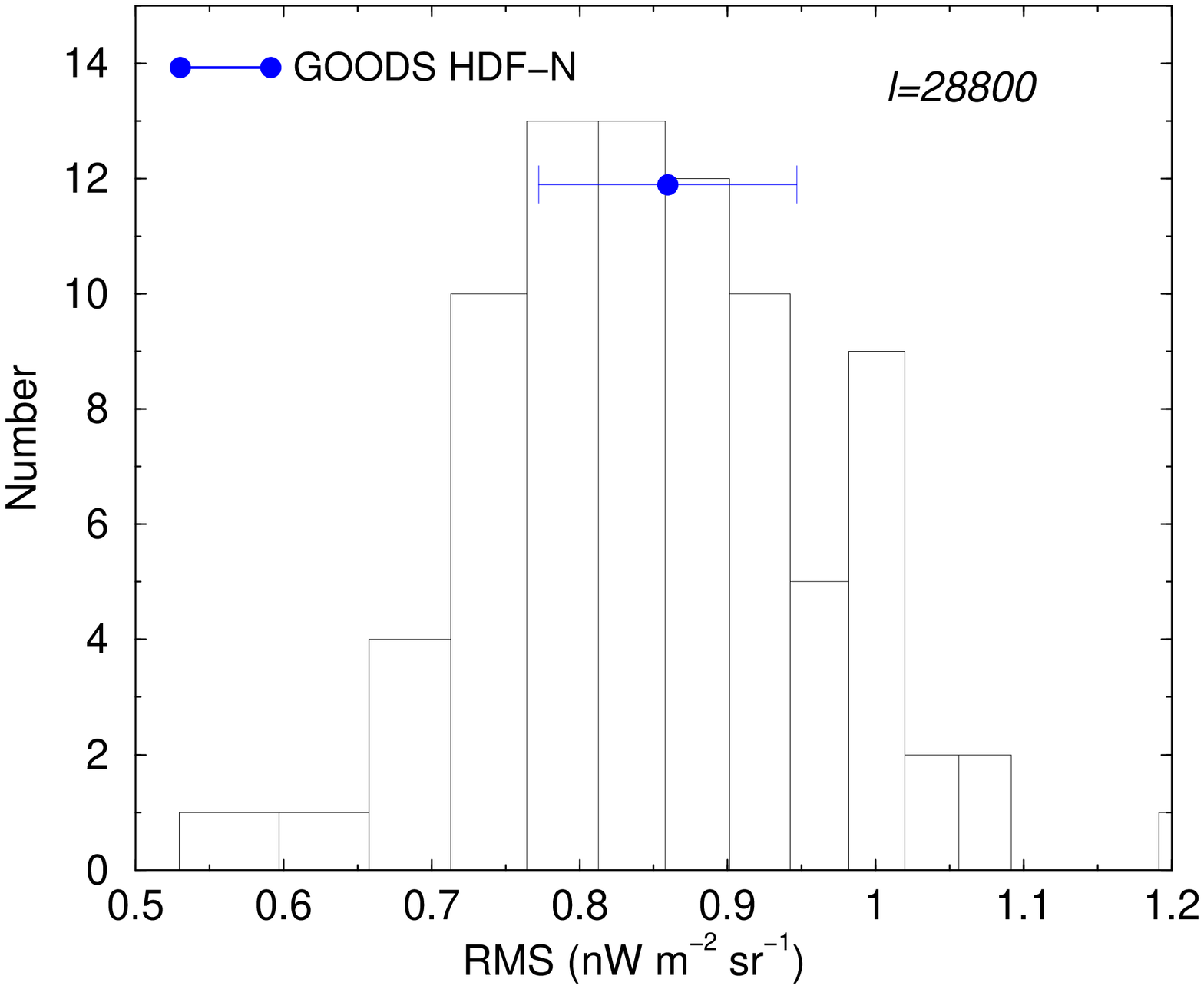,width=2.8in,angle=-90}}    
\caption{
The rms fluctuation in the IRB from resolved counts with magnitudes between 17 and 19 at 3.6$\mu$m at an angular scale of $\sim$ 4.5 
arcminutes ($\ell=7200$; left panel) and  $\sim$ 1.2 arcminutes ($\ell=28800$; right panel) when measured in 100 independent fields of 
$\sim$ 0.05 degrees randomly extracted
from the {\it Spitzer} IRAC Bo\"otes image. For comparison, we also show the rms fluctuation level estimate at these multipoles from GOODS HDF-N and the estimated 1$\sigma$ error bar of the rms based on the boot-strap sampling described in the text. The variance in 100 independent fields within Bo\"otes of GOODS-sized fields 
are consistent with the single error estimated for the GOODS HDF-N catalog directly through bootstrap sampling.
This agreement suggests that we are properly accounting for the cosmic variance of the angular power spectrum.
}      
\label{fig:bootes-irac}      
\end{figure*}            

\begin{figure}[!t]      
\centerline{\psfig{file=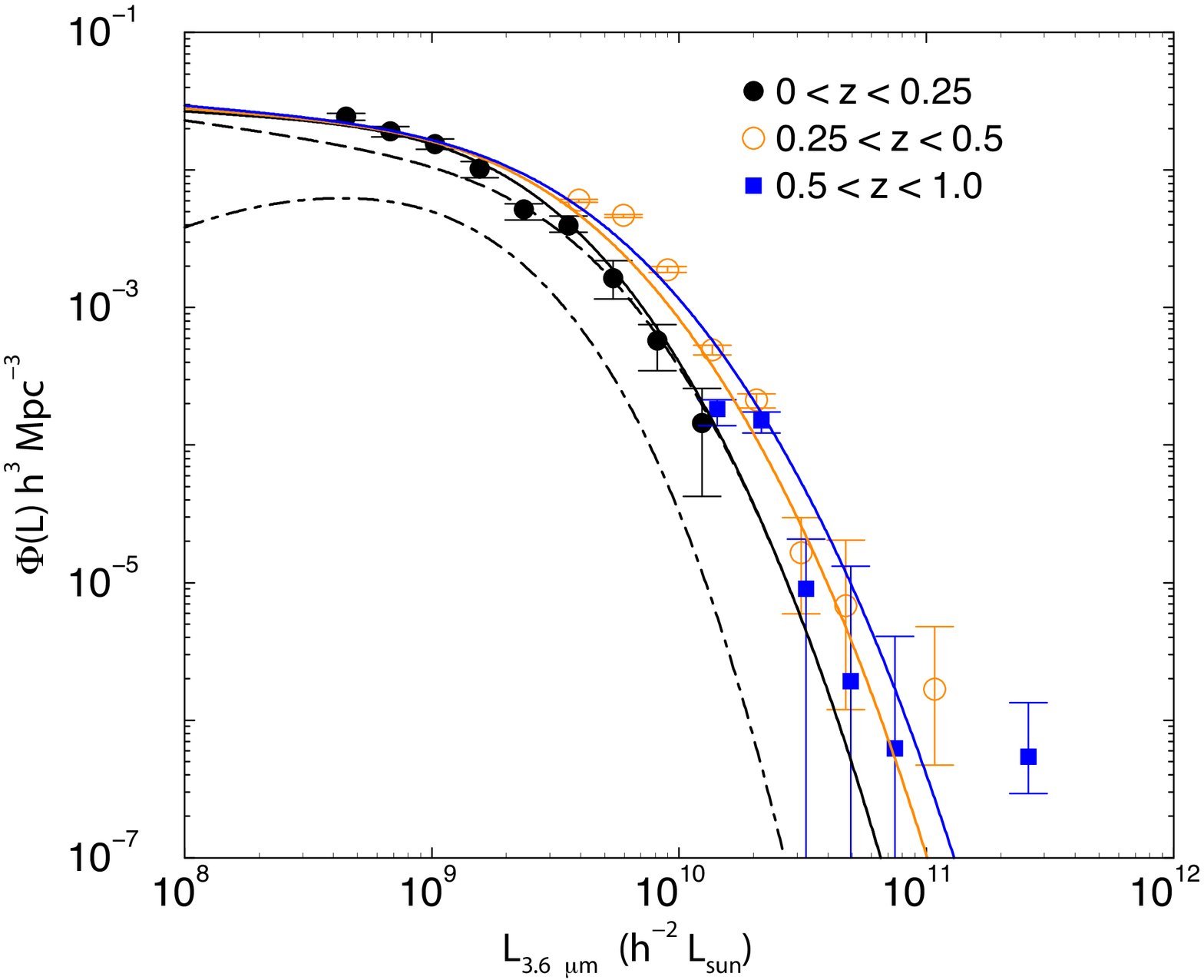,width=3.3in,angle=0}}
\caption{3.6 $\mu$m galaxy luminosity functions (LFs) as a function of redshift. Plotted data points are from Babbedge et al. (2006). The solid curves show the LF based on a conditional luminosity function model. For $0 < z < 0.25$, we also show
the LF from central (dashed line) and satellite (dot-dashed line) separately.
}
\label{fig:lf}      
\end{figure}

In Figure~4, we show only the clustering component as this is directly related to      
the large-scale structure of the source distribution. The measurements 
are binned in multipole. Errors in each bin were
computed through a boot-strap method, where a part of the source distribution was replaced by white noise and the power spectrum      
measured 100 times by repeating the process. The power spectrum error is estimated by the bootstrap sample variance associated with the multiple    
measurements and we have established that these errors are consistent with the theoretically expected error for $\tilde C_l$ of
\begin{equation}
\Delta(\tilde C_l) = \sqrt{\frac{2}{f_{\rm sky}(2l+1)\Delta l}} \left(\tilde C_l + \tilde C_l^{\rm shot-noise}\right) \, ,
\end{equation}
to better than 10\% when $f_{\rm sky}$ is the fraction of sky covered in the survey and $\Delta l$ is the bin width for each
$\tilde C_l$ measurement.

To test how well we have estimated errors in our clustering spectrum, in Figure~5 we plot 
the rms fluctuation in the IRB from resolved counts with magnitudes between 17 and 19 at 3.6$\mu$m at an angular scale of $\sim$ 4.5 
arcminutes ($\ell=7200$; left panel) and  $\sim$ 1.2 arcminutes ($\ell=28800$; right panel) when measured in 100 independent fields of 0.05 degrees randomly extracted
from the {\it Spitzer} IRAC Bo\"otes image. For comparison, we also show the rms fluctuation level estimate at these multipoles from GOODS HDF-N and the estimated 1$\sigma$ error bar of the rms based on the jackknife sampling described in the text. The variance in 100 independent fields within Bo\"otes of GOODS-sized fields 
are consistent with the error estimated from GOODS  data, suggesting that we are properly accounting for the cosmic variance.
This comparison suggests that our method to estimate errors leads to reasonable estimates of the uncertainties in the clustering spectrum.           
      
\begin{table*}[hbt]\small      
\caption{\label{table:data}}      
\begin{center}      
\begin{tabular}{ccccccccc}      
\tableskip\hline\hline\tableskip      
Survey & Band & Area & Mag. Range & N$_{\rm sources}$ & $\lambda$ I$(\lambda)$ & $\int S^2 dN/dS\, dS$ & $[\lambda$ I$(\lambda)]^{c}$ & $[\int S^2 dN/dS\, dS]^{c}$\\      
 &  & [deg.$^2$]& [Vega] &  & [nW m$^{-2}$ sr$^{-1}$] & [nW$^2$ m$^{-4}$ sr$^{-1}$] & [nW m$^{-2}$ sr$^{-1}$] & [nW$^2$ m$^{-4}$ sr$^{-1}$]\\    
\tableskip\hline\tableskip      
NDWFS & $L$ & 6.3 & 17-19 & 2.2 $\times 10^5$ & 1.75  & 3.36 $\times 10^{-8}$ & 2.09 & 3.96 $\times 10^{-8}$ \\      
(Bo\"otes)       & $K_S$ & 3.9 & 17-19 & 6.3 $\times 10^4$ & 2.9 & 2.05 $\times 10^{-7}$ & &\\    
       & $J$ & 3.9 & 17-19 & 2.7 $\times 10^4$ & 5.51 & 1.72 $\times 10^{-6}$ & &\\    
       & $J$ & 3.9 & 19-21 & 6.2 $\times 10^4$ & 2.13 &  1.10 $\times 10^{-7}$ & &\\    
GOODS & $L$ & 0.046 & 17-19 & 2.1 $\times 10^3$  & 1.99  &  3.69 $\times 10^{-8}$ & 2.20 & 4.06 $\times 10^{-8}$\\      
(HDF-N) & $L$ & 0.046 & 19-21 & 5.5 $\times 10^3$  & 0.86 & 2.65 $\times 10^{-9}$ & 1.24 & 3.69 $\times 10^{-9}$\\      
      & $L$ & 0.046 & 21-22.5 & 4.8 $\times 10^3$ & 0.17 & 1.02 $\times 10^{-10}$ & 0.39 & 2.26 $\times 10^{-10}$\\      
\tableskip\hline      
\end{tabular}\\[12pt]      
\begin{minipage}{6.0in}      
NOTES.---%
Survey parameters associated with clustering measurement in both GOODS (HDFN; Dickinson et al. 2003) and NDWFS Bo\"otes (Jannuzi \& Dey 1999)
fields in $J$-, $K_S$, and $L$-bands.  With deep GOODS HDF-N  data we consider three bins in magnitudes in the $L$-band.    
The $L$-band Bo\"otes field data are from {\it Spitzer} while $J$- and $K_S$ data involve catalogs of sources detected with    
ground-based imaging of the Bo\"otes field by the FLAMINGOS Extragalactic Survey (FLAMEX; Elston et al. 2006).   
The last four columns tabulate the mean background intensity and the shot-noise 
for uncorrected and corrected counts (with $[\lambda$ I$(\lambda)]^{c}$ and $[int S^2 dN/dS\, dS]^{c}$), 
respectively (see, Section~3.1 for details on completeness corrections). 
\end{minipage}       
\end{center}      
\end{table*}

In Table~1, we list the magnitude bins considered for clustering measurements with the GOODS HDF-N and the Bo\"otes catalogs    
at $L$-band from {\it Spitzer} IRAC and $J$- and $K_S$-bands from ground-based imaging of the Bo\"otes field.    
The first bin is selected to provide an overlap between the two catalogs. Bo\"otes with the largest sky coverage of the surveys      
allows large-scale linear clustering measurements to be made, while GOODS  allows measurements in the mildly non-linear to      
non-linear regime. In Table~1, we also list the mean IR background produced by sources in each of the bins      
as well as the shot-noise to the anisotropy measurements; the former comes from the source count distributions while      
the latter comes from direct anisotropy measurements at small angular scales.       
     
\section{Analytical model for IR source clustering}      
      
Following Cooray et al. (2004), we consider a simple halo model-based description for the angular clustering of IR sources. The angular power      
spectrum can be written as      
\begin{equation}      
C_l = \int dz \frac{dr}{dz} \frac{a^2(z)}{d_A^2} i^2_\lambda(z)  P_{ss}\left(k=\frac{l}{d_A},z\right) \, ,      
\end{equation}      
where $a(z)=(1+z)^{-1}$ is the scale factor,
$d_A$ is the comoving angular diameter distance and $r$ is the radial distance, while $i_\lambda(z) $ is the       
mean emissivity per comoving volume of IR sources at wavelength $\lambda$ and $P_{ss}(k)$ is the power spectrum of the clustered sources.   
Based on a comparison to the exact formula involving integrals over spherical Bessel functions, 
the Limber approximation (Limber 1954) used here to describe the clustering spectrum of IR sources in the multipole space is
accurate to better than  0.2\% at multipoles less than 10 and is accurate to even higher precision for degree scale fluctuations
at $\ell \sim 10^3$.
    
Instead of modeling the exact relation between IR luminosity and halo mass, we assume that over a given      
magnitude range, we can simply scale source fluctuations by the mean IR background produced by those sources to get back to       
fluctuations in the IR intensity. In this case, the angular power spectrum takes a simpler form of $C_l = \bar{I}^2 w_l$,      
where $\bar{I}^2$ is the cumulative intensity of sources in the magnitude range of interest      
and $w_l$ is the projected angular power spectrum of the sources      
relative to the density field such that $w_l = \int d\rad w^2(z)/d_A^2 P_{ss}\left(k=\frac{l}{d_A},z\right)$, where $w(z)$ is the      
normalized radial distribution of sources such that $\int dz w(z)=1$.      
We make use of the statistical redshift distribution implied by the photometric redshift estimates    
for galaxies in the Bo\"otes catalog  (see, Figure~3). 

In addition to the halo model, clustering can also be predicted based on linear clustering spectrum that ignores growth of
perturbations and non-linearities that are generated at low redshifts as structures grow under gravitational evolution.
For this, we simply take the linear spectrum in our fiducial cosmology, and then calculate
$C_l^{\rm lin} = \bar{I}^2\int d\rad w^2(z)/d_A^2 P^{\rm lin}\left(k=\frac{l}{d_A},z\right)$, where $D^2(z)P^{\rm lin}(k)$ is the linear power spectrum
and $D(z)$ is the linear theory growth function (Peebles 1980).

This linear theory spectrum, however, is not an appropriate description since it fails to describe clustering at small angular scales
or multipoles above a few hundred. Thus, more accurate descriptions of the underlying source spectrum is required.
For this, we make use of the halo model where the  three-dimensional source power spectrum contains two terms 
with $P_{ss}(k) = P^{1h}(k) + P^{2h}(k)$ (see, Cooray \& Sheth 2002). These two terms are      
 clustering between IR galaxies in two different      
halos (2h) and clustering of galaxies within the same halo (1h), and given by      
\begin{eqnarray}      
P^{2h}(k) &=&  \left[\int dM\; n(M)\; b(M) \frac{\langle N_{\rm gal}(M)\rangle}{{\bar n}_g} u(k|M) \right]^2 P^{\rm lin}(k)\nonumber \\      
P^{1h}(k) &=&  \int dM\; n(M)\; \frac{2 \langle N_{\rm s} \rangle \langle N_{\rm c} \rangle u(k|M)  + \langle N_{\rm s} \rangle^2 u^2(k|M)}{\bar{n}_g^2} \, ,      
\label{eqn:pk}      
\end{eqnarray}      
respectively. Here, $u(k|M)$ is the normalized density profile       
in Fourier space (e.g., Navarro, Frenk \& White 1996),       
$n(M)$ is the halo mass function (e.g., PS mass function of Press \& Schechter 1974), $b(M)$ is the halo bias relative to the      
linear density field (e.g., Mo, Jing \& White 1997),  and $\bar{n}_g$ is the number densities of galaxies defined below.
As written, the 2-halo term with $P^{2h}(k)$ traces the linear power spectrum, but at large $k$, through $u(k|M)$, it departs
from the linear theory prediction alone. 

To calculate halo occupation numbers, $ \langle N_{\rm s} \rangle$ and $\langle N_{\rm c} \rangle$ with 
$\langle N_{\rm gal}(M)\rangle=  \langle N_{\rm s} \rangle + \langle N_{\rm c} \rangle$,
we consider two separate ways to model the galaxy distribution. First, we consider
halo occupation numbers where all galaxies are treated the same regardless of their luminosity or flux and then built
a more complete model using conditional luminosity functions (Cooray 2006a) that are normalized to reproduce the
3.6 $\mu$m galaxy LFs measured in {\it Spitzer} SWIRE surveys out to a redshift of 1.0  (Babbedge et al. 2006).
       
\subsection{Simple halo model}

In the simplest description for the halo occupation, 
the source distribution within halos is described analytically as      
$\langle N_{\rm gal}(M) \rangle = 1 + \langle N_{\rm s}(M)\rangle$ when $M > M_{\rm min}$ and 0 otherwise,       
with the assumption of a central galaxy in each halo ($\langle N_{\rm c}(M)\rangle=1$)
 and a power-law distribution of satellites that scales with halo mass as      
$\langle N_{\rm s}(M)\rangle = A(M/M_{\rm min})^\beta$; We take parameters $A,\beta$, and $M_{\rm  min}$ to be free      
parameters to be determined from the data.   Here, $M_{\rm min}$ is the minimum halo mass scale at which galaxies appear 
over the magnitude range of interest; this is not the same as the minimum mass scale at which any galaxy can appear since
smaller halos may host galaxies with lower luminosities, but we do expect an overall cut-off for the appearance of
any galaxy at some low mass scale for dark matter halos (see Section~6.2).  The mean number density of galaxies is      
$\bar{n}_g = \int dM \; n(M) \langle N_{\rm gal}(M) \rangle $.     
    
\subsection{Conditional luminosity functions}

While the above description does not separate galaxies based on their luminosity, 
we can extend our formalism to separate the occupation number to a fixed central galaxy  and satellites around them    
through conditional luminosity function models (CLF; Cooray \& Milosavljevi\'c 2005; Cooray 2006a,b; van den Bosch et al. 
2006). These CLFs describe the average number of galaxies with luminosities between $L$ and $L+dL$ that resides
in halos of mass $m$ at a redshift of $z$, while halo occupation numbers defined above describe the total number of
galaxies in a dark matter halo of mass $m$ regardless of the luminosity of those galaxies.
The CLF models are designed to model not just clustering statistics, but also one-point moments of the
distribution such as LFs and number counts.  We refer the reader to Cooray (2006a) for details, but here provide a summary
and related fitting functions.

We define redshift-dependent CLFs such that
\begin{eqnarray}
\Phi(L|M,z)&=&\Phi^{\rm c}(L|M,z)+\Phi^{\rm s}(L|M,z) \nonumber \\
\Phi^{\rm c}(L|M,z)  &=& \frac{1}{\sqrt{2 \pi} \ln(10)\sigma_{\
rm c} L} \times \nonumber \\
&& \quad \quad \exp \left\{-\frac{\log_{10} [L /L_{\rm c}(M,z)]^2}{2 \sigma_{\rm
 c}}\right\}  \nonumber \\
\Phi^{\rm s}(L|M,z) &=& A(M,z) L^{\gamma(M)} \, .
\label{eqn:clf}
\end{eqnarray}
Given the CLF, the galaxy LF is obtained through
\begin{equation}
\Phi(L,z) = \int dM\, \frac{dn}{dM}(z)\, \Phi(L|M,z) \, .
\end{equation}
Finally, given the information related to observed magnitude ranges,
one can integrate over the corresponding luminosity distribution at a given redshift
to obtain a more accurate estimate of the occupation numbers
\begin{eqnarray}
\langle N_{\rm c}(M,z) \rangle &=& \int dL \, \Phi^{\rm c}(L|M,z) \nonumber \\
\langle N_{\rm s}(M,z) \rangle &=& \int dL \, \Phi^{\rm s}(L|M,z) \, .\nonumber \\
\end{eqnarray}
These can then replace the calculation procedure outlined at the beginning of this Section to
calculate the angular power spectrum. In above, central galaxy CLF takes a log-normal shape
while satellite CLF takes a power-law behavior with luminosity.   Note that the above two integrals
over a log-normal and a power-law functions in luminosity, generally recover the mass dependence of
the simple halo occupation number written in the previous section with 
$\langle N_{\rm gal}(M) \rangle = 1 + \langle N_{\rm s}(M)\rangle$\, where 1 is for the central galaxy with $\langle N_{\rm c}(M,z) \rangle$ and
$\langle N_{\rm s}(M)\rangle$ is a power-law with mass.

An important ingredient in CLF models is the mapping between the central galaxy luminosity $L_{\rm c}$ 
and halo mass (Vale \& Ostriker 2004) as a function of redshift (Cooray 2006b).
This relation is described with a general fitting formula given by
\begin{equation}
L_{\rm c}(M,z) = L_0(1+z)^\alpha \frac{(M/M_1)^{a}}{[b+(M/M_1)^{c(1+z)^\beta}]^{1/d}}\, .
\label{central}
\end{equation}
Following the procedures outlined in Cooray (2006a,b), we first model the LF of 3.6 $\mu$m galaxies 
between redshifts of 0 and 1 (from SWIRE survey; Babbedge et al. 2006). In Figure~6, we show the measured LF and
the model based on CLFs with parameters for the $L_{\rm c}(M,z)$ relation of
$L_0=1.3\times10^{9} L_{\sun}$, $M_1=10^{11} M_{\sun}$,
$a=4.0$, $b=0.7$, $c=3.75$, and $d=0.22$. The redshift dependence is captured by fitting parameters $\alpha$ and
$\beta$  (see, Cooray 2006b) and at 3.6 $\mu$m we have found these to take values of -0.2 and -0.05, respectively,
while the dispersion of central galaxy luminosity (at a fixed mass $M$) in log-luminosity units as written in equation~8
is $\sigma=0.2$.
As discussed in Cooray (2006b) there are large degeneracies between these parameters when fitting to the LF and, to
obtain an overall analytical model to compare with clustering measurements, 
we only consider the best-fit description with LF curves shown in Fig.~6.

Finally, we also need a description of the total luminosity content of a halo, $L_{\rm tot}(M,z)$, and this is done by using the
same fitting function as for central galaxies in equation~\ref{central} with $c=3.5$ while all parameters are kept the same.
The luminosity content of satellites is then $L_{\rm s}(M,z)\equiv L_{\rm tot}(M,z)-L_{\rm c}(M,z)$, and this fixes the
normalization $A(M,z)$ in equation~\ref{eqn:clf} by requiring that 
 $L_{\rm s}(M,z)=\int_{L_{\rm min}}^{L_{\rm max}} \Phi^{\rm sat}(L|M,z)LdL$.
When written the satellite CLF, we take $\gamma$ to be independent of mass with a value of  -1. This is consistent
with optical cluster LFs, but these models can be improved significantly once similar LF measurements of galaxy groups and
clusters are made at {\it Spitzer} bands and not just the field LF that averages over a large distribution of halo masses.

\section{Results \& Discussion}

\subsection{Clustering}

In Figure~4 (left panel), we show the clustering measurements and a model description of the data. GOODS data, due to low sky area of 0.092 deg.$^2$ between the two HDF-N and CDF-S fields,      
measure only non-linear clustering, but between magnitudes 17 and 19 (45 and 7 $\mu$Jy), Bo\"otes data allow the angular power spectrum to be extended to      
large angular scales related to the linear part of the angular power spectrum. The combination, Bo\"otes and GOODS, shows evidence
for the transition between linear to  non-linear clustering with a departure from a single power-law clustering spectrum.
With counts corrected for completeness, over the angular scales of overlap,
we find a good agreement between the clustering spectrum determined in the
large-area Bo\"otes field and the ones determined with GOODS images.
If not completeness corrected, we find a $\sim 10$\% offset between Bo\"otes clustering spectrum and the GOODS clustering spectrum at
multipoles of $10^4$, with Bo\"otes field underestimating power relative to the GOODS measurements. 

In addition to the combination of GOODS and Bo\"otes in the $L$-band,
we also detect a departure from a power-law in the ground-based 
FLAMEX Bo\"otes $J$-band data though $K_S$-band clustering measurements could be fitted with a power-law (Figure~4 right panel). 
In Figure~4, the lines show model descriptions of the angular power spectrum of    
source clustering using the CLF model for 3.6 $\mu$m galaxies in the bin of 17 to 19 magnitudes and 21 to 21.5 magnitudes separately.
These predictions make use of the statistical redshift distribution for $L$-band galaxies estimated based on photometric redshift estimates
shown in Figure~3 (left panel). Our model fits are not strongly sensitive to assumptions about the redshift distribution as long as    
we do not take all sources to be either at very low or very high redshift. 
Note the factor of 100 difference in the y-axis scale between the left and the right panels of Figure~4;     
for sources in the same magnitude bin of 17 to 19 (255 to 40 $\mu$Jy $J$-band, and 107 to 17 $\mu$Jy $K_S$-band),
 $J$-band sources produce a factor of $\sim$ 7.5 larger rms fluctuations, $\sqrt{l^2 C_l/2\pi}$,     
in the resolved IRB relative to sources in the same magnitude ranges in the $L$-band. This large increase is consistent with the difference in the mean intensity of the     IR background between these two different wavelengths and differences in the average clustering bias of $J$-band and $L$-band galaxies.

The CLF model-based descriptions of IR galaxy clustering spectra, as shown in Figure~4, require knowledge on the relation
between both central galaxy luminosity and halo mass and the total luminosity and halo mass. As outlined in 
Section~5.2, we obtain these relations based on model descriptions to the measured LFs at 3.6 $\mu$m in the {\it Spitzer} SWIRE survey
(Babbedge et al. 2006). This procedure is similar to the ones used to described galaxy clustering in SDSS at low redshifts
and DEEP-2 and Lyman-break galaxies at high redshifts (Cooray 2006a; Cooray \& Ouchi 2006). A different modeling technique, but with the same
underlying approach, is available in van den Bosch et al (2006) where CLFs are  a priori assumed to be Schechter-function shapes.
Similarly, we also use the simple halo model (in Section~5.1) to model fit the data as well. Here, the models 
require three parameters, $M_{\rm min}$,  the minimum  dark matter halo mass at which IR galaxies begin to appear,     
$\beta$, the power-law slope of the satellite occupation with halo mass, and $A$, 
the normalization of the satellite occupation number relative to the    
central galaxy occupation of unity in halos with mass above $M_{\rm min}$. Instead of model fits to the CLFs, which involve
a large number of parameters, here we will directly constrain parameters such as  $A$ and $\beta$.
     The right panel of Figure~4 shows clustering measurements at $J$- and $K_S$-bands using the source catalogs from the ground-based imaging data    
of the Bo\"otes field. Again, in addition to clustering measurements, we also show model descriptions 
based on the halo model for galaxies with magnitude brighter than 19 in both $J$- and $K_S$-bands.     We have not
attempted to model fit fainter galaxy samples due to issues related to the completeness of the catalogs.

\begin{figure*}[!t]      
\centerline{\psfig{file=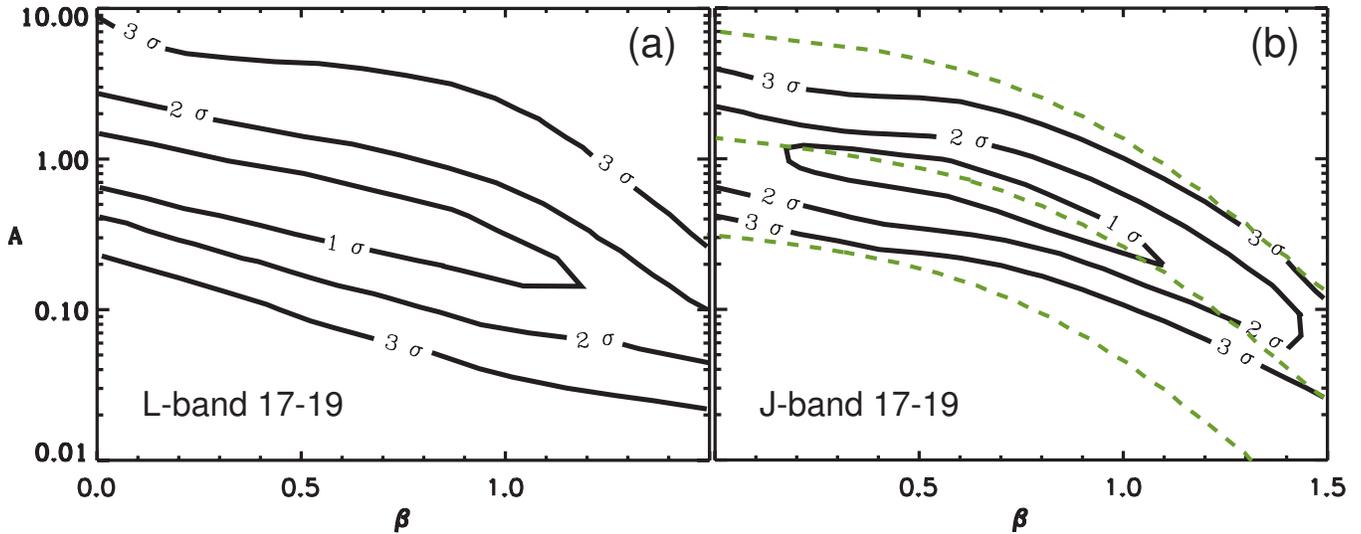,width=7in,angle=0}}      
\caption{(a) Model fits to the $L$-band Bo\"otes and GOODS HDF-N clustering measurements in the magnitude bin of 17 to 19 (45 to 7.0 $\mu$Jy), and    
(b) $J$-band Bo\"otes field clustering in the magnitude bin of 17 to 19 (255 to 40 $\mu$Jy). We show the constraints on parameters      
that describe the satellite occupation number, following Section~5.1 (see text for details).
In (a), for the $L$-band galaxies with magnitudes between 17 and 19,    
model fits suggest $\beta < 0.98$  at the 1$\sigma$ level, while $A$ is between 0.4 and 1.5.  In (b), for $J$-band galaxies with magnitudes between 17 and 19,    
$0.27 <  \beta < 0.89$ at the 1$\sigma$ level, while $A$ is between 0.4 and 1.1.
In (b), the dashed lines    
show constant $\bar{n}_g = \int dz w(z) \int dN/dM \langle N_{\rm gal} \rangle/\int dz w(z)$; the degeneracy direction in $\beta$-$A$ plane traces the number density    
 of IR sources, normalized to the redshift distribution of the sources, $w(z)$.}    
\label{fig:contour}      
\end{figure*}      

In Figure~7, we consider a likelihood model fit to the angular clustering measurements only by taking two parameters,      
$\beta$ involving the slope of the satellite counts as a function of halo mass and $A$ the overall      
normalization of the central-to-satellite galaxy occupation number to be free  while     
$M_{\rm min}$ is taken to be fixed at  $5 \times 10^{11}$ M$_{\sun}$ for galaxies with magnitudes in the range 17 to 19. This $M_{\rm min}$ value
is not the minimum mass for galaxies with any luminosity to appear in halos but rather those in the appropriate magnitude ranges.  Since we
are considering IR bright galaxies in the magnitude range between 17 and 19, this minimum mass is consistent with the expected value  with the
CLF model, though the cut off is not sharp at a fixed mass scale given the dispersion we allow between luminosity and halo mass . 
With large catalogs, the accuracy of clustering measurements      
can be improved and further parameters determined, but at this stage, it is unlikely that more than 2 parameters      
can be extracted from the data. We only model fit the range of 17 to 19 magnitude in both the $L$-band and the $J$-band, since at deeper bins with      
either the GOODS data or ground-based Bo\"otes data alone, the clustering measurements are mostly limited to non-linear scales only leading to      
large degeneracies between parameters in the occupation number, and the photometric estimates for the source distribution of these    
galaxies are less certain than bright galaxies in the magnitude bin between 17 and 19.       
    
Based on the constraints on $A$ and $\beta$ shown in Figure~7,     
we estimate the halo mass scale at which two IR sources with a magnitude between 17 and 19 in the $L$-band appear in the same halo,    
or when $\langle N_{\rm gal}(m) \rangle =2$, is $(0.9-7)\times 10^{12}$ M$_{\sun}$. This is consistent with the number 
in the CLF-based model with a value of $\sim 2\times 10^{12}$ M$_{\sun}$. Note that in CLF-based models, due to the
large number of parameters involved, we do not vary them to fit them individually, but obtain an overall description consistent with
 3.6 $\mu$m LFs between redshifts of 0 and 1 and also use a single set of values for the input parameters when comparing to measurements.
In terms of CLF models, this is the best we can do at this stage.
The mass scale we estimate for galaxies at $z$ of 1 can be compared to previous estimates in the literature. For example,    
using conditional luminosity function based  models of galaxy clustering (Cooray 2006a) compared to DEEP2 (Coil et al. 2004) and COMBO-17 (Phleps et al. 2003)     measurements of galaxy clustering at redshifts between 0.7 and 1.3 suggests that the halo mass scale at which bright 
galaxies with luminosities above $L_\star$ at $z \sim 1$ start to appear as satellites    
is  $\sim (1-8)\times 10^{12} M_{\sun}$ at the 1$\sigma$ confidence range, in good agreement with 
the same scale we have established for $z \sim 1$ IR galaxies. 

The satellite halo occupation number for IR sources in the $L$-band is such that the power-law    
slope with mass is below 0.98 at the 1$\sigma$ confidence level.  For $J$-band galaxies with magnitudes between 17 and 19,    
the power-law slope with mass of the satellite occupation number is $ 0.27 < \beta < 0.89$ at the 1$\sigma$ confidence level,
which is consistent with the expected value for the slope 0.72 with halo mass from out CLFs.
As shown in Figure~7 right panel, we illustrate the degeneracy direction    
in the $\beta$-$A$ parameter plane by plotting lines of constant average density of galaxies calculated as $\bar{n}_g = \int dz w(z) \int dM dN/dM \langle N_{\rm gal}\rangle$,    
where $w(z)$ is the normalized redshift distribution of sources used for clustering measurements. The degeneracy in $\beta$-$A$ is such that one traces constant $\bar{n}_g$ values.     This number density,  as a function of redshift, is directly available with LFs and was the basis
to normalize our CLF description. In future, if LF measurements can be improved (the measurements have low signal-to-noise at $z > 0.5$
as can be seen in Fig.~5) and also be extended to higher redshifts, then we can hope to 
improve constraints on the occupation number, or more importantly on the conditional luminosity function. 
An analysis based on such an approach for $z \sim 3$ to 4 Lyman-break galaxies are available in Cooray \& Ouchi (2006)
and we hope to return to this topic again in a later paper using additional measurements we have now begun to
make from the same set of data as used for the clustering measurements here (Cooray et al., in preparation).

\subsection{Anisotropies in the IRB}

To understand how these fluctuations of resolved sources compare with that of the diffuse IRB,     
in Figure~8 left panel we compare our clustering measurements with the anisotropy power spectrum presented in      
Kashlinsky et al. (2005) for the unresolved component with all sources brighter than 22.5 magnitude (0.28 $\mu$Jy) 
removed.  In the case of these unresolved anisotropies, 
the small scale structure is that of a shot-noise with the power spectrum scaling as $\ell^2$; In fact, it      
is this shot-noise that Kashlinsky et al. (2005) used to determine that they had removed resolved sources down to a magnitude      
limit of 22.5. The flattening of the fluctuations at the two lowest $\ell$ bins is taken to be indicative of an excess component in the IRB,      with that component ascribed to first galaxies.     The predictions shown in lines come from the
CLF model with parameters that reproduce the LF.

The right panel of Figure 8 shows the completeness corrected number counts.  The lines here are expectation
for number counts using the CLF by integrating over LFs as a function of volume element. Note that
at magnitudes below 18, the models slightly under predict the counts despite a reasonable agreement with LFs. This difference, however,
comes from stellar contamination to source counts as our counts are not corrected for stars, though based on the stellar
model of Fazio et al. (2004), we do correct for stars in our clustering measurements.

We show three specific CLF models for counts which are all normalized to LFs at $z < 1$, but with differences at the faint end not probed by the LFs. 
The corresponding clustering spectra associated with those counts are shown in the left panel of Fig.~8.
In Fig.~8 right panel, the black lines involve a model where
we set the minimum mass for appearance of at least one galaxy 
regardless of the luminosity in a dark matter halo to be  $3 \times 10^{10}$ M$_{\sun}$
(or effectively a luminosity with a corresponding magnitude below 30 in $L$-band since
our discussion below is for galaxies brighter than 30th magnitude).
That is, at halo masses below this value galaxies are not expected to form due to a variety of reasons from photoionization during
reionization, preheating of gas from AGNs, and/or expulsion of gas from first generation of supernovae. 
This number is consistent with the value generally quoted in the literature based on
semi-analytical models of galaxy formation (e.g., Kauffmann et al. 1993; Benson et al. 2001).
Furthermore, this value is constrained to be below $3 \times 10^{10}$ M$_{\sun}$ using direct likelihood fits to
clustering measurements in SDSS (Cooray 2006a), though this is for galaxies in the $r'$-band.
With this cut-off in halo mass, number counts peak at $L$-band magnitudes around 24.

To obtain an excess of galaxies at the faint-end, we arbitrarily lower this low mass cut off for galaxy formation and set it at a mass
scale of  $10^{8}$ M$_{\sun}$ and take the central galaxy luminosity-halo mass relation to continue down to such low masses.
This introduces an excess of faint galaxies with an increase in the number counts at magnitudes starting from a magnitude of 22.
If we use these counts down to a magnitude limit of 26, the sources between the magnitude range of 22.5
and 26 produce a shot-noise of $\sim 5 \times 10^{-11}$ nW$^2$ m$^{-4}$ sr$^{-1}$, which is
roughly a factor of 8 higher than the shot-noise seen in the fluctuation measurements ($\sim 7.2 \times 10^{-12}$ nW$^2$ m$^{-4}$ sr$^{-1}$). 
Note that such a large slope at the faint-end
has been suggested based on a study of the one-point probability distribution function
of the unresolved IR intensity (Savage \& Oliver 2005), the large difference between the expected and
measured shot-noise level suggests that
$L$-band source counts at the faint-end do not continue to increase down to magnitude limits of 26 with the same slope 
as the one
suggested with completeness-corrected counts between 19 and 22. The order of magnitude lower shot-noise measured by 
Kashlinsky et al. (2005) suggests that the counts either flatten or turn over  consistent with our basic CLF
model with a minimum mass cut off around  $3 \times 10^{10}$ M$_{\sun}$ for appearance of IR galaxies in dark matter halos.

The integrated background associated with the default description for number counts (in black lines) is
$\sim$ 0.08 nW  m$^{-2}$ sr$^{-1}$ between the magnitude range of 22.5 and 30,
suggesting that if the excess clustering suggested by Kashlinsky et al. (2005) is correct, then these  
sources are unlikely to be a significant fraction of the total IRB intensity at 3.6 microns. If sources were to continue with the large positive 
slope to the faint-end, then the background between magnitude limits of 22.5 and 30 would have been close to 1.2 nW  m$^{-2}$ sr$^{-1}$,
which is an appreciable fraction given that the total background is about 10 nW  m$^{-2}$ sr$^{-1}$ at 3.6 $\mu$m.
Note that the background intensity of $\sim$ 0.08 nW m$^{-2}$ sr$^{-1}$ we suggest for faint sources to describe
unresolved source clustering shot-noise is also below
the level of 1 to 2 nW m$^{-2}$ sr$^{-1}$ background intensity from a population of $z > 5$ galaxies that was considered
to reproduce clustering measurements of Kashlinsky et al. (2005) by Salvaterra et al. (2006). 
Salvaterra et al. (2006) models assume that shot-noise is dominating from $\ell$ of 10$^5$ and only consider the contribution coming from
galaxies at high redshifts.

To emphasize the redshift ranges where the contributions are coming from, in our CLF, we put a low redshift cut at $z \sim 5$.
The expected counts are show in green in the right panel with corresponding clustering spectrum for those galaxies
in the left panel of Figure~8. The clustering level is well below the measured level at two  orders of magnitude below
the amplitude of Kashlinsky et al. (2005). Since the clustering spectrum is proportional to square of the background,
see Section ~5, one can  arbitrarily renormalize the counts, as in the case of
models in Salvaterra et al. (2006), to fit Kashlinsky et al. (2005) data, but then
the observed excess lead to a total intensity of IRB an order of magnitude higher. Similarly, counts are also
increased by an order of magnitude.
This renormaization, however, is not correct since it leads to a large surface density of galaxies
at redshifts more than 5 at a level more than allowed by the expected density of dark matter halos at these
redshifts. This expected high density of galaxies has been already realized in Salvaterra \& Ferrara (2006)
who have argued against a large density of first galaxies that dominate the IR background given constraints on the number counts.
This arbitrary normalization is not an issues for our CLF models  since
the galaxy density is constrained given the connection between dark matter halos and galaxies  they host.
Unlike the simple model in Salvaterra et al. (2006) the amplitude is not a free parameter.

More importantly, Figure~8 left panel also shows that the clustering predicted for galaxies in the magnitude range between 22.5 and 26 is
consistent with measurements of Kashlinsky et al. (2005) for unresolved background and for clustering of resolved galaxies.
Here, non-linear clustering of galaxies within dark matter halos continue to be important out to multipoles of 10$^6$. 
For comparison, in Figure~8 left panel, we also show the clustering measurements in  our faintest bin between 21 and 22.5
in GOODS. The excess seen in Kashlinsky et al. (2005) can be described as simply an extension of resolved component that we have
measured and extending to a larger angular scale than for measurements in Kashlinsky et al. (2005). In combination,
there is a consistent description to the data from our CLFs suggesting that the unresolved component seen by
Kashlinsky et al. (2005)  can be described by galaxy counts with magnitudes between 22.5 and 26 at redshifts below 5 with contributions mostly from 
redshifts 1 to 3, and that there is no need to invoke a large surface density of  galaxies at redshifts above 5.

The limited range of multipoles
probed by Kashlinsky et al. (2005) limits us from making an exact statement given that the data can also be modeled
from a clustering spectrum that trace the linear density field at redshifts above 8 (dot-dashed lines following
Cooray et al. 2004). This diffuse background is taken to be produced by Pop-III stars and their spectra are completely different from
known galaxies given the dominant energy in the Lyman-$\alpha$ line, which is redshifted to near-IR wavelengths today (see, 
Santos et al. 2002). Thus, our modeling based on CLFs described above 
will not contain such a separate component especially if the Pop-III formation happens at $z > 8$.
The 3.5 $\mu$m IRB intensity associated with  models in the top and the bottom curve is $\sim$ 2.5 and 0.3 nW m$^{-2}$ sr$^{-1}$, respectively.    
The angular power spectrum of clustering  peaks at a multipole of $\sim 10^3$, corresponding to the peak of the linear clustering when projected at these     high redshifts.  
If the IRB excess is due to this component,  the the clustering excess should peak at multipoles around $\ell^2$ $\sim 10^3$; 
this is simply a reflection of the    
projection associated with the peak of the linear mass power spectrum. As discussed in Cooray et al. (2004; Bock et al. 2006),
in the H-band, the IRB intensity is roughly a factor of 10 higher given that the Pop-III spectrum is sharply rising towards lower 
wavelengths as the  intensity there 
is dominated by the Lyman-$\alpha$ emission associated with recombinations during the reionization of the Universe.    
The two models shown in Figure~8 left panel bracket two extreme regions in terms of the reionization history, the density of first galaxies, 
their clustering     strength or bias, and the Lyman-$\alpha$ photon production.    
It is clear that one should focus on degree to arcminute angular scales for anisotropy studies in the IRB instead of sub-arcminute angular scales     
considered by Kashlinsky et al. (2005) at multipoles above 10$^4$. As is clear from Figure~8 left panel, such imaging need not be very deep    
since degree-scale shallow images that allow removal of sources down to a magnitude of 21 in the $L$-band 
can be easily used to probe the range suggested by the two extreme models.    

To briefly summarize our discussion so far, the excess clustering power spectrum measured in Kashlinsky et al. (2005)
is consistent with faint galaxies with magnitudes between 22.5 and 26 and predominantly at redshifts between 1 and 3
as extrapolated from measured LFs out to $z \sim 1$. These faint galaxies are unresolved in the deep
{\it Spitzer} images, but not a significant contributor to the total intensity of the IRB. If we were to explain the missing
IRB with a new population of sources, such as Pop-III stars, then the clustering spectrum can be explained, but there is
an overall increase in fluctuations at multipoles of $\sim$ 10$^3$ which is not currently probed by unresolved clustering.

\begin{figure*}[!t]      
\centerline{\psfig{file=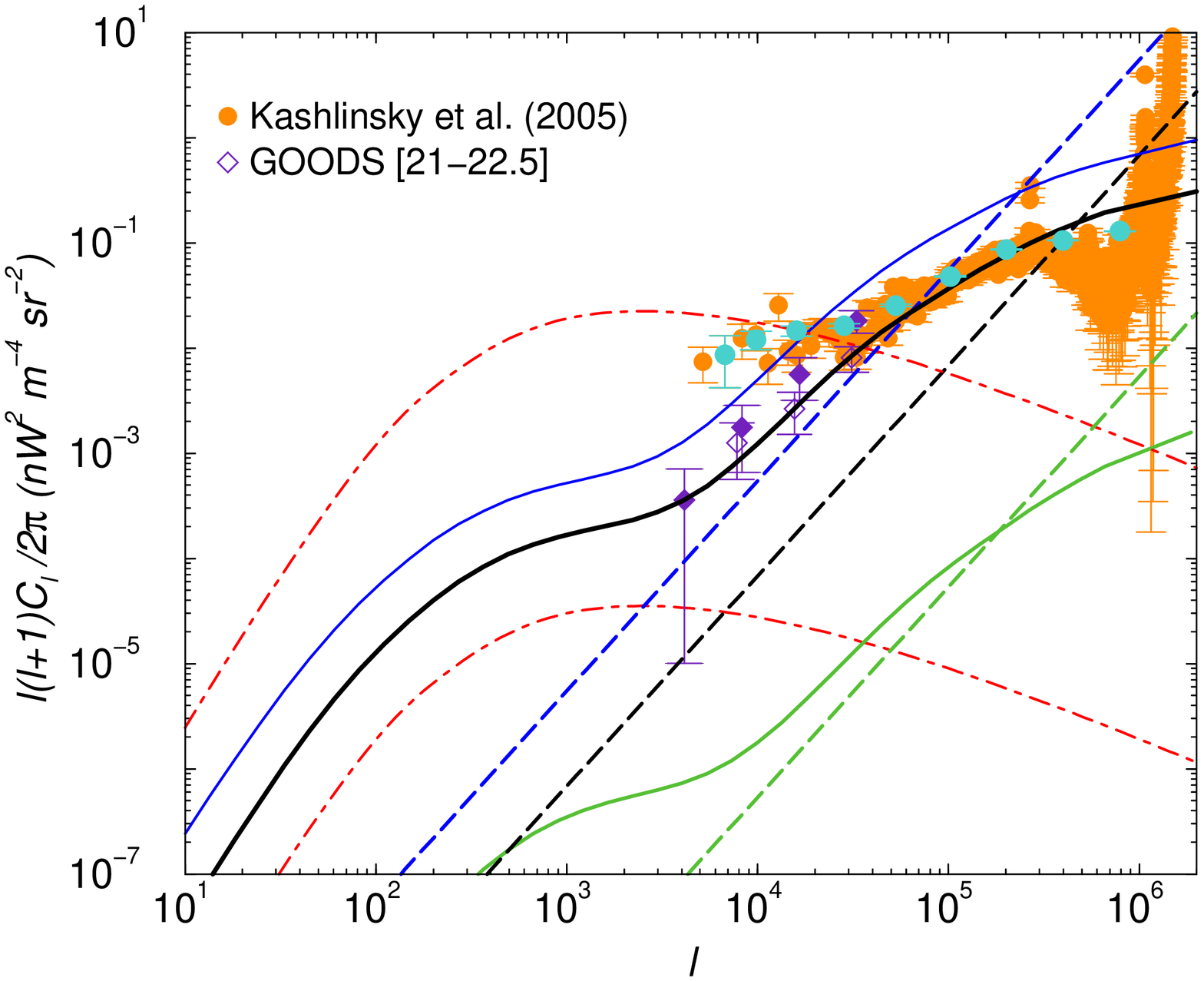,width=2.8in,angle=-90}    
\psfig{file=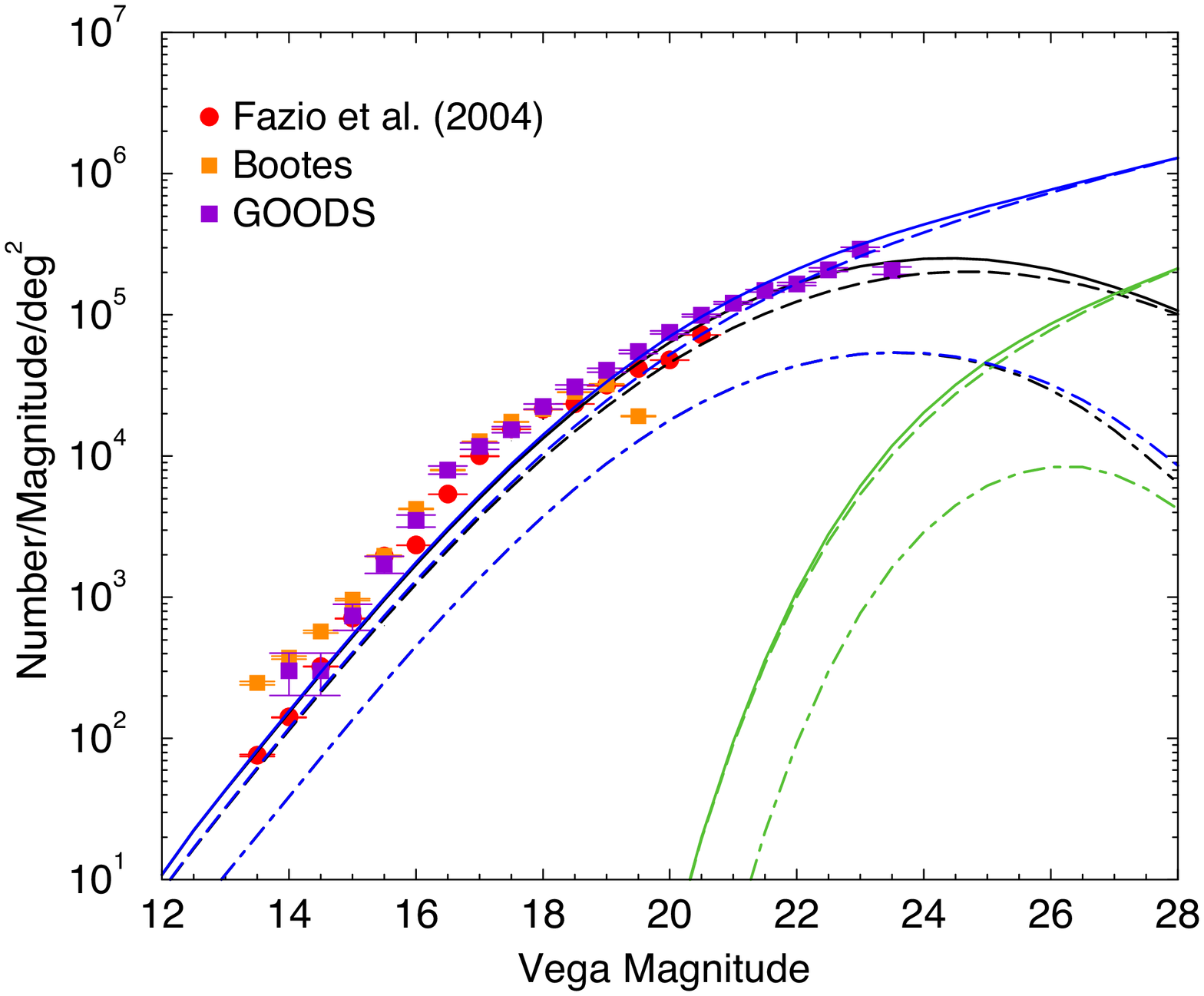,width=3.0in,angle=-90} }
\caption{
IRB fluctuations from a $z> 8$ diffuse component that traces the linear density field (dot-dashed lines),
 galaxy clustering (solid lines), and shot-noise from galaxies (dashed lines that scale as $l^2$ with increasing multipole).
The measurements are from Kashlinsky et al. (2005) in the $L$-band with fainter symbols showing the 
background fluctuations after rebinning the original measurements. We consider
three model descriptions for clustering using the CLF models (see text for details). The corresponding
counts are shown and compared to completeness corrected measured counts in the right panel.
In the right panel, in addition to total counts (solid lines), we also show counts of central (dashed lines) and satellite (dot-dashed lines)
galaxies. These predictions for number counts
are obtained by simply integrating the model LFs over the volume with  luminosity converted to an observed flux as
a function of redshift. The black lines show our default model and the shot-noise component can be roughly described as due to    
point sources with $L$ band magnitudes between 22.5 and 26. This model also reproduces clustering 
suggesting that unresolved clustering spectrum measured by Kashlinsky et al. (2005) can be described by galaxies in the magnitude range between
22.5 and 26. If we populate IR galaxies down to a halo mass of 10$^8$ M$_{\rm sun}$, we find counts shown by the blue line, with a
slope consistent with the faint-end slope suggested in Savage \& Oliver (2005). These counts lead to 
an order of magnitude higher shot-noise than measured with the clustering spectrum of IR anisotropies
and a larger clustering amplitude than measured. Finally, with green lines, we only consider
counts at $z > 5$ for the same model, but in this case, we fail to explain the measurements given that these faint
galaxies are not strongly clustered even at high redshifts.
The two dot-dashed lines show a model expectation for clustering in the diffuse IRB from unresolved sources at $z > 8$ 
prior to complete reionization (Cooray et al. 2004). 
The 3.5 $\mu$m IRB intensity associated with models in the top and the bottom curve is $\sim$ 2.5 and 0.3 nW m$^{-2}$ sr$^{-1}$, respectively.    
The angular power spectrum of clustering  peaks at a multipole of $\sim 10^3$, corresponding to the peak of the linear clustering when 
projected at redshifts greater than 8. }
\label{fig:irb}      
\end{figure*}

\subsection{Wide-Field Surveys for IRB Anisotropy Measurements}

While degree to arcminute angular scale fluctuation measurements allow first-stars at $z>8$  to be more easily identified,    
extending fluctuation measurements to low multipoles is challenging with {\it Spitzer} IRAC images alone      
due to both the small field-of-view and to problems such as uncertainties in relative calibration associated with stitching       
independent fields together to make a larger map. The Bo\"otes field considered here comes      
from a large number of independent images, but we are able to make clustering measurements of resolved sources since      
the catalogs are less affected by issues related to sensitivity variations from field to field. 
Nevertheless, we are also studying the possibility to use the 6 deg.$^2$ central region of the Bo\"otes mosaic to at least place
limits on the amplitude of unresolved source clustering at 10 arcminute angular scales to a degree angular scales.
These results will be presented in an upcoming paper (Cooray et al. 2006).

As mentioned in above, to separate various possibilities, the
 need for large-area clustering measurements is clear from the left panel of Figure~8 (see, also Figure~2 of Bock et al. 2006).
While at multipoles of $10^4$ to 10$^5$ the model
clustering of a $z> 8$ diffuse component overlaps with faint galaxy clustering, at multipoles of $10^3$ the clustering
is well separated. Thus, at degree angular scales, instead of deep images one can use shallow images down to a brighter
magnitude limit to search for an excess component than the magnitude limit one has to image when probing 
the same diffuse component with images that allow clustering studies only at arcminute angular scales and below.
In consideration of the two issues discussed so far, large-angular scale excess and the large difference between
measured IRB intensity and the predicted intensity based on counts 
at lower wavelengths than studied by {\it Spitzer}, we suggest observations that attempt to resolve 
the nature of the IRB excess should concentrate on shallow, wide-field imaging at wavelengths below 3.6 $\mu$m.
      
The Cosmic Infrared Background Explorer (CIBER; Bock et al. 2006) is designed directly to address the ``missing source'' problem in IRB      
that is significant between 1 $\mu$m and 2 $\mu$m through shallow, but wide-field images. 
CIBER uses wide-field imagers with a field-of-view of 4 deg.$^2$ in both I- and H-bands, supported by spectrometers to determine the contribution from zodiacal light and establish the      
spectrum of IRB intensity from 0.7 to 1.8 $\mu$m.      
With the wide-field coverage, CIBER can measure      
clustering of IRB light from multipoles below $10^2$ to $10^4$ covering the linear to non-linear regime from a single image.      
Furthermore, CIBER will image the Bo\"otes field so that existing source counts from {\it Spitzer} and other ground-based imaging      
data can be used to remove point sources. The combination of resolved and unresolved source clustering      
by extending studies such as the one presented here      
should allow us to pin down the sources responsible for the missing IRB and directly address the ``missing source'' problem      
of the IRB.      

In addition to CIBER, at wavelengths above 2.2 $\mu$m, the North Ecliptic Pole wide-field survey with the IR Camera on
{\it Akari} (Matsuhara et al. 2006) will provide an additional dataset to study large-scale fluctuations in the
near-IR background. The shallow survey useful for clustering measurements will span over 6.2 deg.$^2$ and will allow

clustering measurements from the linear to non-linear scales similar to the angular power spectrum established here for
3.6 $\mu$m resolved sources in the Bo\"otes field.  Finally, it could be that after the depletion of cryogen necessary for longer wavelength 
instruments, {\it Spitzer} will continue to operate only the first two channels of IRAC. Such a scenario could be
exploited for a very wide-field survey which in return could be used to further improve the clustering
measurements of both resolved and unresolved components. While the deep imaging data with {\it Spitzer} IRAC of the GOODS HDF-N and CDF-S 
fields have now allowed clustering measurements down to very faint magnitudes, these measurements are restricted to the non-linear regime. It will be very useful to measure clustering for sources at magnitude levels of 21 and fainter so that background anisotropy
measurements at degree angular scales can be properly combined with resolved source clustering at these faint magnitude levels to obtain
a complete picture of the IRB anisotropies.

\section{Summary}    
    
We have presented a measurement of angular power spectrum of the clustering of near-IR  sources    
at 3.6 $\mu$m in {\it Spitzer} imaging data of the GOODS HDF-N, the GOODS CDF-S, and the NDWFS Bo\"otes field in several source magnitude bins.    
We also measured the angular power spectrum of resolved sources in the Bo\"otes field at $K_S$ and $J$-bands    
using ground-based IR imaging data. In the three bands, $J$, $K_S$, and $L$, we  have    
detected the clustering of galaxies on top of the shot-noise power spectrum at multipoles between      
 $\ell \sim 10^2$ and $10^5$. The angular power spectra range from the large, linear scales to small, non-linear scales of galaxy clustering,  and show a clear departure from a power-law clustering spectrum for $L$-band galaxies
when clustering in the Bo\"otes field is combined with GOODS. We consider a halo model      
to describe clustering measurements and  establish the halo occupation number parameters of    
IR bright galaxies at redshifts around unity.   
  
The typical halo mass scale at which    
 two or more IR galaxies with $L$-band magnitude between 17 and 19 are found in the same halo is    
between $9 \times 10^{11}$ M$_{\sun}$ and $7 \times 10^{12}$ M$_{\sun}$ at the 68\% confidence level; 
this is consistent with the previous halo mass estimates  for galaxies at $z\sim1$ from clustering studies in
surveys such as DEEP2. We have also
discussed our results in the context of a recent measurement related to the unresolved IR background   (IRB)    
anisotropies based on {\it Spitzer} imaging data by Kashlinsky et al. (2005).  While the unresolved  IRB fluctuations were measured at sub-arcminute angular scales,    
we have argued that the nature of suggested excess clustering can be best studied with shallow, wide-field images     
that can make measurements of clustering from a few degree to arcminute angular scales with resolved $L$-band 
sources removed down to a magnitude level of about 21 and $J$-band     sources removed down to a magnitude limit of 19.
An attempt at making a wide-field image of the near-IR sky  is planned    
with the Cosmic Infrared Background Explorer (CIBER; Bock et al. 2006) at I- and H-bands, {\it Akari} at K-band and above, 
and potentially with {\it Spitzer} in the $L$-band.

{\it Acknowledgments:}       
This research was carried out at the Jet Propulsion Laboratory, California Institute of Technology, under a contract with the 
National Aeronautics and Space Administration and funded through the Director's Research and Development Fund program.
We thank FLAMEX survey, and the survey PI Anthony Gonzalez,  
 for making catalogs of the Bo\"otes field $J$- and $K_S$-bands sources publicly available.  
We thank Thomas Babbedge for providing us with electronic files of the SWIRE luminosity functions.
This work is based on  observations made with the {\it Spitzer Space Telescope}, 
which is operated by the Jet Propulsion Laboratory, California
Institute of Technology, under NASA contract 1407.        
We thank an anonymous referee for a helpful report that improved the presentation of measurements and results in this paper.
The numerical code related to CLF models described here is available from http://www.cooray.org/lumfunc.html .

\end{document}